\documentstyle[prd,preprint,aps,epsfig,floats]{revtex} 

\def\Journal#1#2#3#4{{#1} {\bf #2} (#4) #3}

\def\NPB{{\em Nucl. Phys.} B}
\def\PLB{{\em Phys. Lett.}  B}
\def\PRL{\em Phys. Rev. Lett.}
\def\PRD{{\em Phys. Rev.} D}

\def\PTP{\em Prog.~Theor.~Phys.}

\def\ra{\rightarrow}
\def \etv{E_T\!\!\!\!\!\!/~~} 

\begin{document}

\draft


\title{Phenomenology of Higgs bosons in the Zee-Model}

\author{
Shinya Kanemura\protect\( ^{1,2}\protect \)\thanks{
Email: kanemu@particle.physik.uni-karlsruhe.de
},
Takashi Kasai\protect\( ^{3,4}\protect \)\thanks{
Email: kasai@post.kek.jp
},
Guey-Lin Lin\protect\( ^{5}\protect \)\thanks{
Email: glin@cc.nctu.edu.tw
},\\
Yasuhiro Okada\protect\( ^{3,6}\protect \)\thanks{
Email: yasuhiro.okada@kek.jp
},
Jie-Jun Tseng\protect\( ^{5}\protect \)\thanks{
Email: u8627512@cc.nctu.edu.tw
},
and {}
C.-P. Yuan\protect\( ^{2,7}\protect \)\thanks{
Email: yuan@pa.msu.edu
}}

\address{
$^1$~\emph{Institut f{\"u}r Theoretische Physik, Universit{\"a}t Karlsruhe, D-76128 Karlsruhe, 
Germany}\\
$^2$~\emph{Physics and Astronomy Department, Michigan State University, 
           East Lansing, MI 48824-1116, USA}\\
$^3$~\emph{Theory Group, KEK, Tsukuba, Ibaraki, 305-0801 Japan}\\
$^4$~\emph{Department of Accelerator Science,
The Graduate University for Advanced Studies,
Tsukuba, Ibaraki, 305-0801 Japan}\\
$^5$~\emph{National Chiao Tung University, Hsinchu 300, Taiwan }\\
$^6$~\emph{Department of Particle and Nuclear Physics, 
The Graduate University for Advanced Studies,
Tsukuba, Ibaraki, 305-0801 Japan}\\
$^7$~\emph{Theory Division, CERN, CH-1211, Geneva, Switzerland} 
}


\maketitle
\thispagestyle{empty}
\vspace*{3cm}
\begin{center}
{\it To appear in Physical Review D}  
\end{center}

\newpage 

\begin{abstract}
To generate small neutrino masses radiatively, the Zee-model introduces 
two Higgs doublets and one weak-singlet charged Higgs boson to its 
Higgs sector. From analyzing the renormalization group equations, we 
determine the possible range of the lightest CP-even Higgs boson ($h$) 
mass and the Higgs boson self-couplings as a function of the cut-off scale 
beyond which either some of the coupling constants are strong enough to 
invalidate the perturbative analysis or the stability of the electroweak 
vacuum is no longer guaranteed. 
Using the results obtained from the above analysis, we find that the 
singlet charged Higgs boson can significantly modify the partial
decay width of $h \ra \gamma \gamma$ via radiative corrections, and its
collider phenomenology can also be drastically different from that of the 
charged Higgs bosons in the usual two-Higgs-doublet models.

\end{abstract}

\setcounter{footnote}{0}
\renewcommand{\thefootnote}{\arabic{footnote}}

\section{Introduction}

\label{sec:intro}
From the atmospheric and solar neutrino data, there is increasing evidence 
for neutrino oscillations\cite{neuosc}.
If this is a correct interpretation, the Standard Model (SM) has to be 
extended to incorporate the small masses of the neutrinos suggested by
data. There have been several ideas proposed in literature 
to generate small neutrino masses. The Zee-model is one of such 
attempts\cite{zee,zeeref,Jarlskog_et_al,Jarlskog_may00,Smirnov_Tanimoto}. 
In this model, all flavor neutrinos are massless at the tree level, and 
their small masses are induced radiatively through one-loop diagrams. 
For such a mass-generation mechanism to work, it is necessary to 
extend the Higgs sector of the SM to contain at least two weak-doublet 
fields and one weak-singlet charged scalar field.
Although some studies have been done to examine the interaction 
of the leptons and the Higgs bosons in the 
Zee-model\cite{ng}, 
the scalar (Higgs) sector of the model remains unexplored in detail.
In this paper we study the Higgs sector of the Zee-model to clarify 
its impact on the Higgs search experiments, either at the CERN LEP-II, 
the Run-II of the Fermilab Tevatron, the CERN Large Hadron Collider 
(LHC), or future linear colliders (LC's).

Experimental search for the Higgs boson has been continued at the 
CERN LEP and the Fermilab Tevatron experiments. 
In the LEP-II experiments, the Higgs boson with the mass less than about 
110 GeV has been excluded if its production cross section and decay modes 
are similar to those of the SM Higgs boson\cite{lep2h}. 
Run-II of the Tevatron can be sensitive to a SM-like Higgs
boson with the mass up to about 180 GeV, provided that the integrated 
luminosity of the collider is large enough 
(about $30 \,{\rm fb}^{-1}$)\cite{run2h}.
Furthermore, the primary goal of the CERN LHC experiments is to guarantee 
the discovery of a SM-like Higgs boson for its mass as large as about 
1 TeV\cite{lhch}, which is the upper bound of the SM Higgs boson mass. 
(For a Higgs boson mass beyond this value, the SM is no longer a 
consistent low energy theory.)

When the Higgs boson is discovered, its mass and various decay properties 
will be measured to test the SM and to distinguish models of new physics 
at high energy scales.
For example, the allowed mass range of the lightest CP-even Higgs boson ($h$) 
can be determined by demanding the considered theory to be
a valid effective theory all the way up to some cut-off 
energy scale (\( \Lambda  \)).
For \( \Lambda =10^{19} \) GeV (i.e., the Planck scale), 
the lower and upper bounds of the 
SM Higgs boson masses are  137 GeV and 175 GeV, respectively \cite{smh19}.
The Higgs mass bounds for the two-Higgs-doublet-model (THDM) were also 
investigated\cite{chivkura,kko2hdm} with and without including the 
soft-breaking term with respect to the discrete symmetry that protects the 
natural flavor conservation.
It was found in Ref.~\cite{kko2hdm} that the lower bound of 
the lightest CP-even Higgs boson is about 100 GeV 
in the decoupling regime where only one neutral
Higgs boson is light as compared to the 
other physical states of Higgs bosons.

The Higgs sector of the Zee-model is similar to that of the THDM except for 
the existence of an additional weak-singlet charged Higgs field, so that 
the physical scalar bosons include two CP-even, one CP-odd and 
two pairs of charged Higgs bosons.
In this paper, we shall first determine the upper and lower bounds 
for the lightest CP-even Higgs boson mass ($m_h$) as a function of the 
cut-off scale $\Lambda$ of the Zee-model, using renormalization group 
equations (RGE's).\footnote{%
For the model with see-saw mechanism for neutrino mass generation the
Higgs mass bound has been studied as a function of cut-off scale 
in Ref.~\cite{see-saw_Higgs}.
}
We show that the upper and lower mass bounds for $h$ are almost the 
same as those in the THDM. 
We also study the possible range of the Higgs-boson self-coupling 
constants at the electroweak scale as a function of $\Lambda$. 
By using these results, we examine effects of the additional loop 
contribution of the singlet charged Higgs boson to the 
partial decay width of \( h\rightarrow \gamma \gamma  \).  
We show that, by taking $\Lambda=10^{19}$ GeV, 
the deviation of the decay width from the SM prediction can be about 
{$-$20\% } or nearly {$+$10\%} for $m_{h}$ between 125 GeV and 140 GeV 
when the mass of the isospin singlet charged Higgs boson is taken to be 
around 100 GeV.
The magnitude of the deviation becomes larger for lower cutoff scales 
and smaller masses of the singlet charged Higgs boson. 
If we choose $\Lambda=10^{4}$ GeV and the singlet charged Higgs boson 
mass to be 100 GeV, the positive deviation can be greater than $+$30\% 
($+$40\%) for $m_{h^{}}=125$ GeV ($140$ GeV). 
Such a deviation from the SM prediction could be tested at the LHC,   
the $e^+e^-$ LC and the $\gamma\gamma$ option of 
LC\cite{test_hgaga_lhc,test_e+e-lc,test_hgaga_lc}. 
We also discuss phenomenology of the singlet charged Higgs 
boson at present and future collider experiments, which is found 
to be completely different from that of 
the ordinary THDM-like charged Higgs bosons. 
To detect such a charged Higgs boson at LEP-II experiments, 
experimentalists have to search for their data sample with 
$e^\pm$ or $\mu^\pm$ plus missing energy, in contrast to the 
usual detection channels: either $\tau \nu$ or $cs$ decay modes.   

This paper is organized as follows. 
In Sec.~\ref{sec:zeemodel}, we introduce the Higgs sector of 
the Zee-model and review the neutrino masses and mixings 
in this model which are consistent with the atmospheric and solar neutrino 
observations. Numerical
results on the possible range of the mass and coupling constants 
of the Higgs bosons are given in
Sec.~\ref{sec:RGE_analysis}. 
In Sec~\ref{sec:h-2gamma}, we discuss the one-loop effect 
of the extra-Higgs bosons in the Zee-model to the partial 
decay width of \( h\rightarrow \gamma \gamma  \)  and its 
impacts on the neutral Higgs-boson search at high-energy colliders. 
The phenomenology of the charged Higgs boson that comes from   
the additional singlet field is discussed in Sec.~\ref{sec:charged}. 
In Sec.~\ref{sec:conclusion}, we present additional discussions 
and conclusion. 
Relevant RGE's for the Zee-model are given in the Appendix. 

\section{Zee-model}

\label{sec:zeemodel}

To generate small neutrino mass radiatively, the Zee-model
contains a \( SU(2)_{L} \) singlet charged scalar field 
\( \omega^{-} \), in addition to two \( SU(2)_{L} \)
doublet fields \( \phi _{1}\), and \(\phi _{2} \). 
The  Zee-model Lagrangian is written as:
\begin{equation}
{\cal L}={\cal L}_{kin}+{\cal L}_{ll\omega }+{\cal L}_{Yukawa}-V(\phi_{1},\phi_{2},\omega^{-}) \, ,
\end{equation}
 where
\begin{eqnarray}
{\cal L}_{kin} & = & \left| D_{\mu }\phi _{1}\right| ^{2}+
\left| D_{\mu }\phi _{2}\right| ^{2}+
\left| D_{\mu }\omega^{-}\right| ^{2}
    +i\overline{q_{_{L}}}\gamma ^{\mu }D_{\mu }q_{_{L}}+
 i\overline{u_{_{R}}}\gamma ^{\mu }D_{\mu }u_{_{R}}+
 i\overline{d_{_{R}}}\gamma ^{\mu }D_{\mu }d_{_{R}}\nonumber \\
 & &  +i\overline{l_{_{L}}}\gamma ^{\mu }D_{\mu }l_{_{L}}+
 i\overline{e_{_{R}}}\gamma ^{\mu }D_{\mu }e_{_{R}} 
    +\sum _{a=SU(3),SU(2),U(1)}\frac{1}{4}F_{\mu \nu }^{a^{2}} \, ,
\end{eqnarray}
\begin{eqnarray}
{\cal L}_{ll\omega }=f_{ij}\overline{l_{i_{L}}}(i\tau
_{2})(l_{j_{L}})^{C}\omega ^{-}+f_{ij}\overline{l_{i_{L}}}^{C}(i\tau
_{2})l_{j_{L}}\omega ^{+} ,
\label{fij_int}   
\end{eqnarray}
where $i,j$ ($=1,2,3)$ are the generation indices, and 
\begin{eqnarray}
V(\phi _{1},\phi _{2},\omega^-) & = & m_{1}^{2}\left| \phi _{1}\right| ^{2}+
m_{2}^{2}\left| \phi _{2}\right| ^{2} 
 +m_{0}^{2}\left| \omega^{-}\right| ^{2}\nonumber \\
 &  & -m^{2}_{3}(\phi _{1}^{\dagger }\phi _{2}+
\phi ^{\dagger }_{2}\phi _{1})
- \mu \widetilde{\phi _{1}}^{T}i\tau _{2}\widetilde{\phi _{2}}\,\omega^{-}
+ \mu \phi_2^{T} i \tau_2 \phi_1\, \omega^+     \nonumber \\
 &  & +\frac{1}{2}\lambda _{1}\left| \phi _{1}\right| ^{4}+
 \frac{1}{2}\lambda _{2}\left| \phi _{2}\right| ^{4}+
 \lambda _{3}\left| \phi _{1}\right| ^{2}\left| \phi _{2}\right| ^{2}
 \nonumber \\
 &  & +\lambda _{4}\left| \phi ^{\dagger }_{1}\phi _{2}\right| ^{2}+
 \frac{\lambda _{5}}{2}\left[ \left( \phi ^{\dagger }_{1}\phi _{2}\right) ^{2}+
 \left( \phi ^{\dagger }_{2}\phi _{1}\right) ^{2}\right] \nonumber \\
 &  & +\sigma _{1}\left| \omega^{-}\right| ^{2}\left| \phi _{1}\right| ^{2}+
 \sigma _{2}\left| \omega^{-}\right| ^{2}\left| \phi _{2}\right| ^{2}+
 \frac{1}{4}\sigma _{3}\left| \omega^{-}\right| ^{4} \, .
 \label{Higgs_potential} 
\end{eqnarray}
In the above equations, 
 $q_L$ is the 
left-handed quark doublet with an implicit generation index
while $u_R$ and $d_R$ denote the right-handed singlet quarks. 
Similarly,  $l_L$ and $e_R$ denote the left-handed and right-handed 
leptons in three generations. 
The charge conjugation of a fermion field is defined as 
\( \psi ^{C}\equiv C\overline{\psi }^{T} \), where
 \( C \) is
the charge conjugation matrix 
(\( C^{-1}\gamma ^{\mu }C=-\gamma ^{\mu T } \)) with
the super index $T$ indicating the transpose of a matrix.
Also,  
\( \phi _{m}=\left( \begin{array}{c}
\phi _{m}^{0}\\
\phi _{m}^{-}
\end{array}\right)\)
and 
\(\widetilde{\phi _{m}}\equiv \left( i\tau _{2}\right) \phi _{m}^{*} \)
with \( m=1,2  \). 
Without loss of generality, we have taken the anti-symmetric
matrix \( f_{ij} \) and the coupling 
\( \mu  \) to be real in the equations (\ref{fij_int}) and 
(\ref{Higgs_potential}). In order to suppress
flavor changing neutral current (FCNC) at the tree level, 
a discrete symmetry, with $\phi_1 \ra \phi_1$, $\phi_2 \ra -\phi_2$,
$\omega^+ \ra + \omega^+$, is imposed to the Higgs sector of the 
Lagrangian, which is only broken softly by the \( m_{3}^{2} \) term 
and the $\mu$ term.  
Under the discrete symmetry there are two possible
Yukawa-interactions;
that is, for type-I
\begin{equation}
{\mathcal L}_{Yukawa-I}  =  
\overline{d_{_{R_{i}}}}\left( y_{_{D}}V_{CKM}^{\dagger }\right)_{ij}\widetilde{\phi _{2}}^{\dagger }q_{_{L_{j}}}
 +\overline{u_{_{R_{i}}}}\left( y_{_{U}}\right) _{ii}\phi _{2}^{\dagger }q_{_{L_{i}}}
 +\overline{e_{_{R_{i}}}}\left( y_{_{E}}\right) _{ii}\widetilde{\phi _{2}}^{\dagger }l_{_{L_{i}}}+h.c. \,,
\end{equation}
and for type-II,
\begin{equation}
{\mathcal L}_{Yukawa-II}  =  \overline{d_{_{R_{i}}}}\left( y_{_{D}}V_{CKM}^{\dagger }\right) _{ij}\widetilde{\phi _{1}}^{\dagger }q_{_{L_{j}}}
 +\overline{u_{_{R_{i}}}}\left( y_{_{U}}\right) _{ii}\phi _{2}^{\dagger }q_{_{L_{i}}}
 +\overline{e_{_{R_{i}}}}\left( y_{_{E}}\right) _{ii}\widetilde{\phi _{1}}^{\dagger }l_{_{L_{i}}}+h.c. \,,
\end{equation}
where \( y_{_{U}} \),\( y_{_{D}} \),\( y_{_{E}} \) are diagonal Yukawa matrices
and \( V_{CKM} \) is the Cabibbo-Kobayashi-Maskawa (CKM) matrix. 
Later, we shall only keep the top Yukawa coupling constants 
\( y_{t}=\left( y_{_{U}}\right) _{33} \)
in our numerical evaluation of the RGE's\footnote{ 
Our analyses will thus be valid in the cases where the effect 
of the bottom Yukawa coupling is sufficiently small; i.e. 
in the region of not too large $\tan\beta$.}.  
In that case, there is no difference 
between the Yukawa couplings of the type-I and the type-II models.
Finally, for simplicity, we assume that 
all \( \lambda _{i} \) and \( m_{i}^{2} \) are real parameters.

Let us now discuss the Higgs sector. 
The \( SU(2)_{L}\times U(1)_{Y} \) symmetry is broken to \( U(1)_{em} \) 
by $\langle \phi_1 \rangle$ and $\langle \phi_2 \rangle$, 
the vacuum expectation values of \( \phi _{1} \) and \( \phi _{2} \). 
(They are assumed to be real so that there is no spontaneous CP violation.)
The number of physical Higgs bosons are two CP-even Higgs bosons 
(\( H \),\( h \)), one CP-odd Higgs boson (\( A \)) and two
pairs of charged Higgs boson (\(S_{1}^\pm\), \( S_{2}^\pm \)). 
We take a convention
of \( m_{H}>m_{h} \) and \( m_{S_{1}}>m_{S_{2}} \). In the basis where two
Higgs doublets are rotated by the angle \( \beta  \), 
with \( \tan \beta =\frac{\left\langle \phi _{2}^{0}\right\rangle }
{\left\langle \phi _{1}^{0}\right\rangle } \),
the mass matrices for the physical states of Higgs bosons are given by 
{\small 
\begin{equation}
\label{neutralmassmatrix}
\! \! M_{N}^{2}\! =\! \! \left[ \! \! \! \begin{array}{lr}
\left( \lambda _{1}\cos ^{4}\beta +\! \lambda _{2}\sin ^{4}\beta +
\! \frac{\lambda }{2}\sin ^{2}2\beta \right) v^{2} & 
\! \! \! \! \! \! \! \! \left( \lambda _{2}\sin ^{2}\beta -
\! \lambda _{1}\cos ^{2}\beta +\! \lambda \cos 2\beta \right) 
\frac{\sin 2\beta }{2}v^{2}\\
\left( \lambda _{2}\sin ^{2}\beta -\! \lambda _{1}\cos ^{2}\beta +
\! \lambda \cos 2\beta \right) \frac{\sin 2\beta }{2}v^{2} & 
\! \! \! \! \! \! \! \! M^{2}+\left( \lambda _{1}+
\lambda _{2}-2\lambda \right) \frac{\sin ^{2}2\beta }{4}v^{2}
\end{array}\! \! \right] \,,
\end{equation}
}for CP-even Higgs bosons,
\begin{equation}
M_{A}^{2}=M^{2}-\lambda _{5}v^{2},
\end{equation}
for CP-odd Higgs boson, and 
\begin{equation}
\label{chargedmassmatrix}
M_{S}^{2}=\left[ \begin{array}{cc}
M^{2}-\frac{\lambda _{4}+\lambda _{5}}{2}v^{2} & -\frac{\mu v}{\sqrt{2}}\\
-\frac{\mu v}{\sqrt{2}} & m_{0}^{2}+\left( 
\frac{\sigma _{1}}{2}\cos ^{2}\beta +
\frac{\sigma _{2}}{2}\sin ^{2}\beta \right) v^{2}
\end{array}\right] \,,
\end{equation}
for charged Higgs bosons. 
Here, \( \lambda \equiv \lambda _{3}+\lambda _{4}+\lambda _{5} \)
and \( M^{2}\equiv m_{3}^{2}/\sin \beta \cos \beta  \). 
The vacuum expectation value $v$ ($\sim 246$ GeV) is equal to 
$\sqrt{2} \sqrt{\langle \phi^0_1 \rangle^2 + \langle \phi^0_2 \rangle^2}$.  
Mass eigenstates for the CP-even and the charged Higgs bosons are 
obtained by diagonalizing the mass
matrices (\ref{neutralmassmatrix}) and (\ref{chargedmassmatrix}), respectively.
The original Higgs boson fields, 
\( \phi _{1} \), \( \phi _{2} \), \( \omega^{-} \),
can be expressed in terms of the physical states 
and the Nambu-Goldstone modes ($G^0$ and $G^\pm$) 
as
\begin{eqnarray}
\phi _{1}^{0} & = & \frac{1}{\sqrt{2}}\left( v\cos \beta +
H\cos \alpha -h\sin \alpha +i(G^{0}\cos \beta -A\sin \beta )\right), \\
\phi _{1}^{-} & = & G^{-}\cos \beta -
(S_{1}^{-}\cos \chi -S_{2}^{-}\sin \chi )\sin \beta, \\
\phi _{2}^{0} & = & \frac{1}{\sqrt{2}}\left( v\sin \beta +
H\sin \alpha +h\cos \alpha +i(G^{0}\sin \beta +A\cos \beta )\right), \\
\phi _{2}^{-} & = & G^{-}\sin \beta +(S_{1}^{-}\cos \chi -
S_{2}^{-}\sin \chi )\cos \beta, \\
\omega^{-} & = & S_{1}^{-}\sin \chi +S_{2}^{-}\cos \chi \, ,
\end{eqnarray}
where the angle \( \alpha  \) and \( \chi  \) are defined from the matrices
which diagonalize the  
\( 2\times 2 \) matrices \( M_{N}^{2} \) and \( M_{S}^{2} \),
respectively. Namely, we have 
\begin{eqnarray}
\left( \begin{array}{cc}
\cos (\alpha-\beta)  & \sin (\alpha-\beta) \\
-\sin (\alpha-\beta)  & \cos (\alpha-\beta) 
\end{array}\right) M_{N}^{2}\left( \begin{array}{cc}
\cos (\alpha-\beta)  & -\sin (\alpha-\beta) \\
\sin (\alpha-\beta)  & \cos (\alpha-\beta) 
\end{array}\right)  & = & \left( \begin{array}{cc}
m_{H}^{2} & 0\\
0 & m_{h}^{2}
\end{array}\right) \,,\\
\left( \begin{array}{cc}
\cos \chi  & \sin \chi \\
-\sin \chi  & \cos \chi 
\end{array}\right) M_{S}^{2}\left( \begin{array}{cc}
\cos \chi  & -\sin \chi \\
\sin \chi  & \cos \chi 
\end{array}\right)  & = & \left( \begin{array}{cc}
m_{S_{1}}^{2} & 0\\
0 & m_{S_{2}}^{2}
\end{array}\right) \, ,
\end{eqnarray}
where \( m_{H}^{2}>m_{h}^{2} \) and \( m_{S_{1}}^{2}>m_{S_{2}}^{2} \). 
The mixing angles $\alpha$ and $\chi$ then satisfy
\begin{eqnarray}
\tan 2\alpha &=& 
\frac{M^2 - \left(\lambda_3 + \lambda_4 + \lambda_5 \right) v^2 }
     {M^2  
      - (\lambda_1 \cos^2 \beta- \lambda_2 \sin^2 \beta) 
        \frac{v^2}{\cos 2 \beta}} \; \tan 2 \beta ,\\  
\tan 2\chi &=& \frac{-\sqrt{2} \mu v }
       {M^2 - m_0^2 - 
   \left( \lambda_4 + \lambda_5 + 
          \sigma_1 \cos^2 \beta + \sigma_2 \sin^2 \beta \right) \frac{v^2}{2}}
     \;,
\end{eqnarray}
which show that $\alpha$ and $\chi$ approaches to 
$\beta - \frac{\pi}{2}$ and zero, respectively\footnote{
Recall that we assumed $m_H^{} > m_h^{}$.}, 
when $M^2$ is much greater than $v^2$, $\mu^2$ and $m_0^2$; i.e., 
in the decoupling regime. In this limit, the massive Higgs bosons from 
the extra weak-doublet are very heavy due to the large $M$ so that 
they are decoupled from the low energy observable.

\begin{figure}
{\par\centering \resizebox*{0.3\textwidth}
{!}{\includegraphics{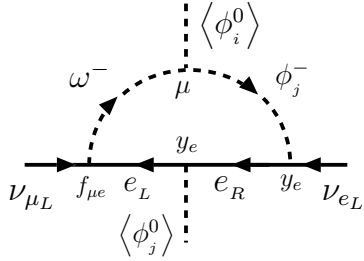}} \par}
\caption{\label{fig:neutrino_mass}
A representative diagram that generates the neutrino mass. 
For type-I, $i=1$, $j=2$, and for type-II, $i=2$, $j=1$.}
\end{figure}

Although neutrinos in this model are massless at the tree level, 
the loop diagrams involving charged Higgs bosons, 
as shown in Fig.~\ref{fig:neutrino_mass},
can generate the Majorana mass terms for all three-flavors of neutrinos.
It was shown \cite{zee} that at the one-loop order, the neutrino mass 
matrix, defined in the basis where the charged lepton Yukawa-coupling 
constants are diagonal in the lepton flavor space, is real and symmetric 
with vanishing diagonal elements.
More explicitly, we have
\begin{equation}
\label{neutrino3x3}
M_{\nu }=\left( \begin{array}{ccc}
0 & m_{12} & m_{13}\\
m_{12} & 0 & m_{23}\\
m_{13} & m_{23} & 0
\end{array}\right) \, ,
\end{equation}
with
\begin{equation}
\label{nmass2fij}
m_{ij}=f_{ij}(m_{e_{j}}^{2}-m_{e_{i}}^{2})\mu \cot \beta 
\frac{1}{16\pi ^{2}}\frac{1}{m_{S_{1}}^{2}-m_{S_{2}}^{2}}
\ln \frac{m_{S_{1}}^{2}}{m_{S_{2}}^{2}} \, ,
\end{equation}
where \( m_{e_{i}} \) \( (i=1,2,3) \) is the charged lepton mass for type-I.
For type-II, $\cot \beta$ should be replaced by $\tan \beta$. 
Note that Eq.~(\ref{nmass2fij}) is valid for $m_{S_i}^{} \gg m_{e_j}$.

The phenomenological analysis of the above mass matrix was given 
in Ref.~\cite{Jarlskog_et_al,Jarlskog_may00}. 
It was concluded that, in the Zee-model, the bi-maximal mixing solution is 
the only possibility to reconcile the atmospheric and the solar neutrino 
data. Here we give a brief summary of their results, for completeness. 
Let us denote the three eigenvalues for 
the neutrino mass matrix, cf. Eq.~(\ref{neutrino3x3}), 
as \( m_{\nu _{1}} \), \( m_{\nu _{2}} \)
and \( m_{\nu _{3}} \), which satisfy 
\( m_{\nu _{1}}+m_{\nu _{2}}+m_{\nu _{3}}=0\).
The possible pattern of the neutrino mass spectrum which is allowed 
in the Zee-model is \( \left| m_{\nu _{1}}\right| 
\simeq \left| m_{\nu _{2}}\right| \gg \left| m_{\nu _{3}}\right|  \),
with \( m_{\nu _{1}}^{2}-m_{\nu _{3}}^{2}\simeq 
m_{\nu _{2}}^{2}-m_{\nu _{3}}^{2}=\Delta m_{atm}^{2} \),
and 
\( \left|m_{\nu _{1}}^{2}-m_{\nu _{2}}^{2}\right| =\Delta m_{solar}^{2} \),
where 
\( \Delta m_{atm}^{2}=O(10^{-3}) \) eV$^{2}$ 
from the atmospheric neutrino data, and 
\( \Delta m_{solar}^{2}=O(10^{-5}) \) eV$^{2}$ 
(MSW large angle solution) or \( O(10^{-10}) \) eV$^{2}$ 
(vacuum oscillation solution) from the solar neutrino data\footnote{
Due to the structure of the mass matrix, cf. Eq.~(\ref{neutrino3x3}), 
only the hierarchy pattern $|m_{\nu_1}| \simeq |m_{\nu_2}| \gg |m_{\nu_3}|$, 
rather than $|m_{\nu_1}| \simeq |m_{\nu_2}| \ll |m_{\nu_3}|$, is realized 
in the Zee-model\cite{Jarlskog_et_al,Jarlskog_may00}}. 
Thus, we have $|m_{\nu_1}| \simeq |m_{\nu_2}| 
                           \simeq \sqrt{\Delta m^2_{atm}}$ 
($m_{\nu_1} \simeq - m_{\nu_2}$) and 
$|m_{\nu_3}| \simeq \frac{\Delta m^2_{solar}}{2\sqrt{\Delta m^2_{atm}}}$. 
The approximate form of the neutrino mass matrix is given by
\begin{equation}
\label{bimaximal_neutrino_mass}
M_{\nu }=\left( \begin{array}{ccc}
0 & \pm \sqrt{\frac{\left| m_{\nu _{1}}m_{\nu _{2}}\right| }{2}} &
 \mp \sqrt{\frac{\left| m_{\nu _{1}}m_{\nu _{2}}\right| }{2}}\\
\pm \sqrt{\frac{\left| m_{\nu _{1}}m_{\nu _{2}}\right| }{2}} & 0 &
 -m_{\nu _{1}}-m_{\nu _{2}}\\
\mp \sqrt{\frac{\left| m_{\nu _{1}}m_{\nu _{2}}\right| }{2}} & 
-m_{\nu _{1}}-m_{\nu _{2}} & 0
\end{array}\right) \, ,
\end{equation}
where the upper (lower) sign corresponds to \( m_{\nu _{1}}<0 \) (\( >0 \))
case, and the corresponding Maki-Nakagawa-Sakata (MNS) matrix\cite{MNS}  
which diagonalizes the neutrino mass matrix is 
\begin{equation}
U=\left( \begin{array}{ccc}
\sqrt{\frac{\left| m_{\nu _{2}}\right| }{\left| m_{\nu _{1}}\right| +
\left| m_{\nu _{2}}\right| }} & \sqrt{\frac{\left| m_{\nu _{1}}\right| }
{\left| m_{\nu _{1}}\right| +\left| m_{\nu _{2}}\right| }} & 0\\
-\frac{1}{\sqrt{2}}\sqrt{\frac{\left| m_{\nu _{1}}\right| }
{\left| m_{\nu _{1}}\right| +\left| m_{\nu _{2}}\right| }} & 
\frac{1}{\sqrt{2}}\sqrt{\frac{\left| m_{\nu _{2}}\right| }
{\left| m_{\nu _{1}}\right| +\left| m_{\nu _{2}}\right| }} & 
\frac{1}{\sqrt{2}}\\
\frac{1}{\sqrt{2}}\sqrt{\frac{\left| m_{\nu _{1}}\right| }
{\left| m_{\nu _{1}}\right| +\left| m_{\nu _{2}}\right| }} & 
-\frac{1}{\sqrt{2}}\sqrt{\frac{\left| m_{\nu _{2}}\right| }
{\left| m_{\nu _{1}}\right| +\left| m_{\nu _{2}}\right| }} & 
\frac{1}{\sqrt{2}}
\end{array}\right) \, ,
\end{equation}
In the above equations, we took the limiting case where \( U_{13}=0 \) and 
\( U_{3 2}=U_{2 3}=\frac{1}{\sqrt{2}} \)~\footnote{
This limit corresponds to $\theta_2=\frac{\pi}{4}$ and   
$\theta_3=0$ in the notation of Ref.~\cite{MNS}.}.
 From Eqs.~(\ref{nmass2fij}) and (\ref{bimaximal_neutrino_mass}), we obtain
\begin{eqnarray}
\left| \frac{f_{12}}{f_{13}}\right|  & \simeq  & 
\frac{m_{\tau }^{2}}{m_{\mu }^{2}} \simeq 3 \times 10^2, \label{eq21}\\
\left| \frac{f_{13}}{f_{23}}\right|  & \simeq  & 
\frac{\sqrt{2}\Delta m_{atm}^{2}}{\Delta m_{solar}^{2}} 
\simeq \left\{ 
\begin{array}{l}
  10^2\;\; ({\rm for \; the \; MSW \; large \; angle \; solution})\\ 
  10^7\;\; ({\rm for \; the \; vacuum \; oscillation \; solution}) 
\end{array} \right. .
\label{eq22}
\end{eqnarray}
Therefore, the magnitudes of the three coupling constants should satisfy the
relation \( \left| f_{12}\right| \gg \left| f_{13}\right| \gg 
\left| f_{23}\right|  \).  This hierarchy among the  couplings $f_{ij}$ 
is crucial for our later discussion on the phenomenology of the 
singlet charged Higgs bosons.

For a given value of the parameters $m_{S_1}$, $m_{S_2}$, $\tan \beta$
and $\mu$, the coupling constants $f_{ij}$ can be calculated from 
Eq.~(\ref{nmass2fij}).
For example, for $m_{S_1}=500$\,GeV, $m_{S_2}=100$\,GeV, $\tan \beta=1$, 
$\mu=100$\,GeV and $m_{12}= 3 \times 10^{-2}$\,eV, 
we obtain $|f_{12}| \sim 3 \times 10^{-4}$. 
As in this example, when $S_1^-$ is rather heavy and the lighter charged 
Higgs boson $S^-_2$ is almost a weak singlet, i.e. 
the mixing angle $\chi$ approaches to zero, it is unlikely that there are 
observable effects to the low energy data\cite{ng};
e.g., the muon life-time, the universality of tau decay into electron or 
muon, the rare decay of $\mu \to e \gamma$, the universality of 
$W$-boson decay into electron, muon or tau, 
and the decay width of $Z$ boson.
When $|f_{ij}|$ are small, we do not expect a large rate in the 
lepton flavor violation decay of a light neutral Higgs boson, such as 
$h \ra \mu^\pm e^\mp$ (the largest one), $h \ra e^\pm \tau^\mp$, or 
$h \ra \mu^\pm \tau^\mp$ (the smallest one).
On the contrary, as to be discussed in Section IV, 
the decay width of $h \ra \gamma \gamma$ can significantly 
deviate from the SM value.

Finally, the phenomenological constraints on \( f_{12} \) were derived in 
Ref.~\cite{Smirnov_Tanimoto}.
From the consistency of the muon decay rate and electroweak precision
test it was found that 
\begin{equation}
\frac{f_{12}^{2}}{\overline{M}^{2}}<7\times 10^{-4}G_{F} \, , \label{f12value}
\end{equation}
where $G_F$ is the Fermi constant, and 
\begin{equation}
\frac{1}{\overline{M}^{2}}=\frac{\sin ^{2}\chi }
{m_{S_{1}}^{2}}+\frac{\cos ^{2}\chi }{m_{S_{2}}^{2}}.
\end{equation}
This means that the $f_{ij}$ cannot be \( O(1) \) unless
the charged Higgs boson masses are at the order of 10 TeV.

\section{Higgs boson mass and couplings through RGE's}

\label{sec:RGE_analysis}
In this section, we determine the bounds on the mass of the
lightest CP-even Higgs boson as a function of the cut-off scale 
of the Zee-model by analyzing the set of renormalization group 
equations (RGE's).
We also study the allowed ranges of the coupling constants, 
especially \( \sigma _{1} \) and \( \sigma _{2} \) 
in Eq.~(\ref{Higgs_potential}). In Sec. IV, they will be used 
to evaluate how much the partial decay width of 
$h \to \gamma\gamma$ can deviate from its SM value  
due to the one-loop contribution from the singlet charged Higgs boson.


The mass bounds are determined in the following manner. 
For each set of parameters
defined at the electroweak scale, the running coupling 
constants are calculated
numerically through RGE's
at the one-loop level. We
require that all the dimensionless coupling constants do
not blow up below a given cut-off scale \( \Lambda  \),  
and the coupling constants satisfy the vacuum stability condition. 
We vary the input parameters at the electroweak scale 
and determine the possible range of the lightest CP-even Higgs 
boson mass as a function of \( \Lambda  \).
In a similar manner, we also study the allowed ranges of various Higgs 
boson self-coupling constants at the electroweak scale as well as
a function of the lightest CP-even Higgs boson mass.

We derived the one-loop RGE's for the Zee-model, and listed them
in the Appendix  for reference. 
For simplicity, 
in the RGE's, we have neglected all the Yukawa coupling constants
(\( y_{u} \), \( y_{d} \), \( y_{e} \)) but the top Yukawa coupling
 \( y_{t} \).\footnote{  
In the model with the type-II Yukawa interaction, 
the bottom-quark Yukawa interaction 
can become important for a large $\tan\beta$. }
Although we kept the new coupling constants 
\( f_{ij} \) in the RGE's listed in the Appendix, 
 we have neglected \( f_{ij} \) in the numerical calculation. This is 
because the magnitudes of these coupling constants are numerically 
too small to affect the final results unless the singlet-charged 
scalar-boson mass is larger than a few TeV [ cf. Eq.~(\ref{f12value}) ]. 
The dimensionless coupling constants relevant
to our numerical analysis are the three
gauge-coupling constants \( g_{1} \), \( g_{2} \), \( g_{3} \), 
the top Yukawa-coupling constant \( y_{t} \), and eight scalar self-coupling 
constants, \( \lambda _{i} \) (\( i=1-5 \)) and
 \( \sigma _{i} \) (\( i=1-3 \)). There are five dimensionful
parameters in the Higgs potential, namely \( m_{1}^{2} \), \( m_{2}^{2} \),
\( m_{3}^{2} \), \( m_{0}^{2} \) and \( \mu  \). Instead of \( m_{1}^{2} \),
\( m_{2}^{2} \), \( m_{3}^{2} \), we take \( v \), \( \tan \beta  \), and
\( M^{2}\equiv m_{3}^{2}/\sin \beta \cos \beta  \),
 as independent parameters, where $v$ ($\sim 246$ GeV) characterizes the
 weak scale and $M$ the soft-breaking scale of the discrete symmetry.

In the actual numerical calculation we first fix \( \tan \beta  \) and \( M \).
For a given mass (\( m_{h} \)) of the lightest CP-even Higgs boson, we solve
one of the \( \lambda _{i} \), which is chosen to be $\lambda_3$ here, 
in terms of other \( \lambda _{i} \).
We then numerically evaluate all dimensionless coupling constants according
to the RGE's. From \( m_{h} \) to \( M \) we use the SM RGE's,
which are matched to the Zee-model RGE's at the soft-breaking scale \( M \). 
\footnote{
The parameter $m_0$ and $\mu$ are only relevant to the charged scalar
mass matrix. In principle, our numerical results also depend on these
parameters through the renormalization of various coupling constants 
from the scale of $m_h$ to the charged scalar mass.
Since these effects are expected to be small,
we calculate the RGE's as if all the scalar-bosons except $h$ decouple
at the scale $M$.
}

We requires the following
two conditions to be satisfied for each scale \( Q \) up to a 
given cut-off scale \( \Lambda  \). 
\begin{enumerate}
\item Applicability of the perturbation theory implies

\begin{equation}
\lambda_{i} (Q) < 8 \pi , 
\quad 
\sigma_{i} (Q) < 8 \pi ,
\quad 
y_{t}^{2} (Q) < 4 \pi \, .
\end{equation}

\item The vacuum stability conditions must be satisfied.   
The requirement that quartic coupling   terms of the scalar potential do not have a 
negative coefficient in any direction leads to the following conditions at 
each renormalization scale \( Q \):
\begin{enumerate}

\item \

\begin{equation}
\lambda _{1}(Q)>0,\quad \lambda _{2}(Q)>0,\quad \sigma _{3}(Q)>0 \, .
\end{equation}

\item \

\begin{equation}
\sigma _{1}(Q)+\sqrt{\frac{\lambda _{1}(Q)\: \sigma _{3}(Q)}{2}}>0 \, ,
\end{equation}
\begin{equation}
\sigma _{2}(Q)+\sqrt{\frac{\lambda _{2}(Q)\: \sigma _{3}(Q)}{2}}>0 \, ,
\end{equation}
\begin{equation}
\overline{\lambda }(Q)+\sqrt{\lambda _{1}(Q)\: \lambda _{2}(Q)}>0 \, ,
\end{equation}
 where \( \overline{\lambda }(Q)=\lambda _{3}(Q)+\min \left( 0,\: 
 \lambda _{4}(Q)+\lambda _{5}(Q),\: 
 \lambda _{4}(Q)-\lambda _{5}(Q)\right)  \).
\item If \( \sigma _{1}(Q)<0 \) and \( \sigma _{2}(Q)<0 \), 
then {\footnotesize 
\begin{equation}
\hspace {-50pt}\overline{\lambda }(Q)+\frac{2}{\sigma _{3}(Q)}
\left\{ \sqrt{\left( \frac{\lambda _{1}(Q)\, 
\sigma _{3}(Q)}{2}-\sigma _{1}^{2}(Q)\right) 
\left( \frac{\lambda _{2}(Q)\, \sigma _{3}(Q)}{2}-
\sigma _{2}^{2}(Q)\right) }-\sigma _{1}(Q)\, \sigma _{2}(Q)\right\} >0
\, .
\end{equation}
}If \( \sigma _{1}(Q)<0 \) and \( \overline{\lambda }(Q)<0 \), 
then{\footnotesize 
\begin{equation}
\hspace {-50pt}\sigma _{2}(Q)+\frac{1}{\lambda _{1}(Q)}
\left\{ \sqrt{\left( \lambda _{1}(Q)\, \lambda _{2}(Q)-
\overline{\lambda }^{2}(Q)\right) \left( 
\frac{\lambda _{1}(Q)\, \sigma _{3}(Q)}{2}-\sigma _{1}^{2}(Q)\right) }-
\sigma _{1}(Q)\, \overline{\lambda }(Q)\right\} >0 \, .
\end{equation}
}If \( \sigma _{2}(Q)<0 \) and \( \overline{\lambda }(Q)<0 \), 
then{\footnotesize 
\begin{equation}
\hspace {-50pt}\sigma _{1}(Q)+\frac{1}{\lambda _{2}(Q)}
\left\{ \sqrt{\left( \lambda _{1}(Q)\, \lambda _{2}(Q)-
\overline{\lambda }^{2}(Q)\right) \left( 
\frac{\lambda _{2}(Q)\, \sigma _{3}(Q)}{2}-\sigma _{2}^{2}(Q)\right) }-
\sigma _{2}(Q)\, \overline{\lambda }(Q)\right\} >0 \, .
\end{equation}
}[ When \( \sigma _{1}(Q) \), \( \sigma _{2}(Q) \) and 
\( \overline{\lambda }(Q) \)
are all negative, the above three conditions are equivalent. ]
\end{enumerate}
\end{enumerate}
In addition to the above conditions, 
we also demand local stability of the potential
at the electroweak scale, namely, we calculate the mass spectrum 
of all scalar fields at the extremum of the potential and 
demand that all eigenvalues of the squared scalar mass are positive. 
We scan the remaining 
seven-dimensional space of \( \lambda _{i} \)
and \( \sigma _{i} \) and examine whether a given mass of the 
lightest CP-even Higgs boson is allowed under the above conditions. 
In this way we obtain the allowed range of $m_h$ as a function of 
\( \tan \beta  \) and \( M \),  for each value
of the cut-off scale \( \Lambda  \).

\begin{figure}[t]
{\par\centering \includegraphics{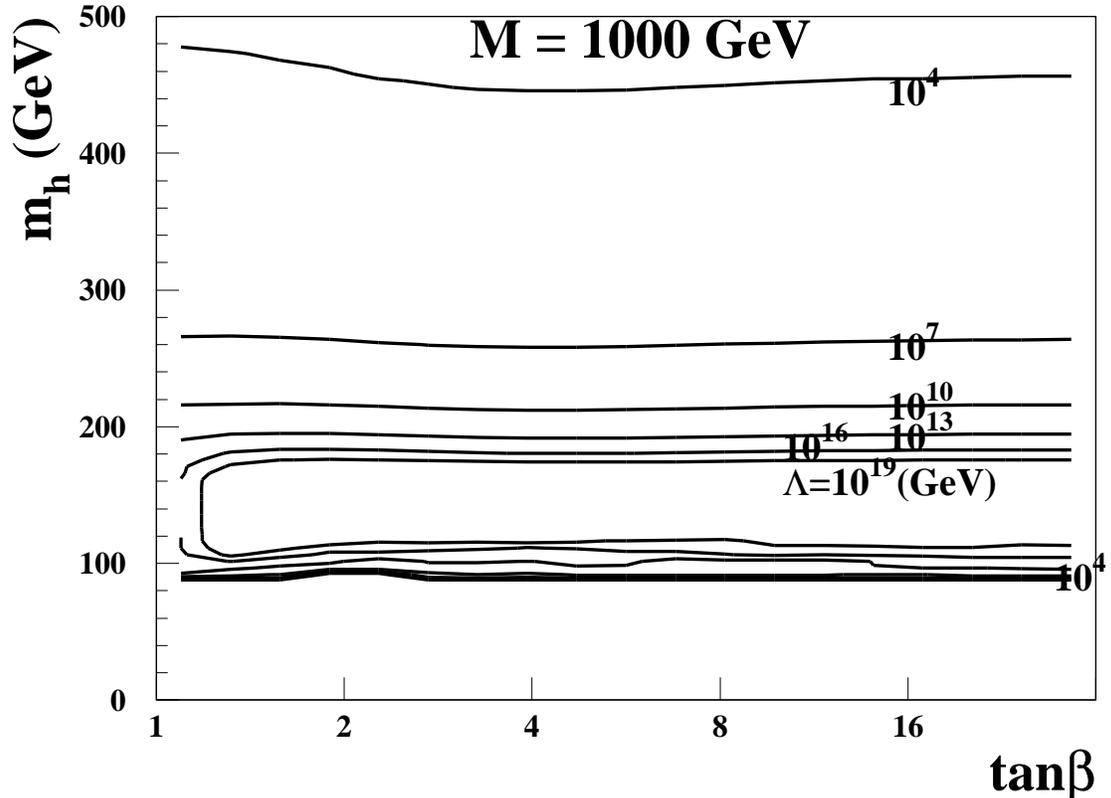} \par}
\caption{
\label{fig:z1000}
The allowed mass range of the lightest 
CP-even Higgs boson for 
\protect\( M=1000\protect \) GeV. $\Lambda$ is the cut-off scale.}
\end{figure}

First, we discuss our result in the decoupling case, in which 
the soft-breaking scale $M$ is much larger than the electroweak scale
$\sim v$, and the masses of all the Higgs bosons but $h$ (and $S_2$) 
are at the order of $M\;$\footnote{ 
In the decoupling regime ($M \to \infty$, which leads to 
$\alpha \to \beta - \frac{\pi}{2}$ and $\chi \to 0$), 
the masses of $h$ and $S_2$ are 
dominated by the $(11)$ component of the mass matrix 
in Eq.~(\ref{neutralmassmatrix}) and the $(22)$ component 
of that in Eq.~(\ref{chargedmassmatrix}), respectively.    
The mass of $h$ is determined by the self-coupling constants $\lambda_i$, 
while that of $S_2$ depends not only on the self-couplings constants 
$\sigma_i$ but also on the free mass parameter $m_0$.
As noticed in the footnote 7, from $m_h$ to $M$, the SM RGE 
are used in our analysis, even if the mass of $S_2$ is smaller than 
$M$. The effect of $S_2$ on the mass bound of $h$ is expected 
to be small, because at the one-loop level the primary effect 
is through the running of $g_1$, whose contribution to the right-handed
side of the RGE for the Higgs-self coupling constant is small.}.
In Fig.~\ref{fig:z1000}, 
the allowed range of \( m_{h} \) is shown as a function of \( \tan \beta  \)
for \( M=1000 \) GeV. 
(We take the pole mass of top quark \( m_{t}=175 \) GeV, 
 \( \alpha _{s}(m_{Z})=0.118 \)
for numerical calculation.) The allowed ranges are shown as contours for six
different values of \( \Lambda  \), i.e. \( \Lambda =10^{19} \),
\( 10^{16} \),\( 10^{13} \),\( 10^{10} \),\( 10^{7} \)
and \( 10^{4} \) GeV. For most values of \( \tan \beta  \), except for small
\( \tan \beta  \) region, the upper bound of \( m_{h} \) is about \( 175 \)
GeV and the lower bound is between 
\( 110 \) GeV and \( 120 \) GeV for the cut-off scale $\Lambda$ to be  
near the Planck scale. The numerical values in this 
figure are very close
to those in the corresponding figure for the THDM 
discussed in Ref.~\cite{kko2hdm}.
Compared to the corresponding lower mass bound in the SM, 
which is \( 145 \) GeV when using the one-loop RGE's, 
the lower mass bound in this model 
is reduced by about \( 30 \) GeV to \( 40 \) GeV. The
reason is similar to the THDM case: 
the lightest CP-even Higgs boson mass is essentially determined 
by the value of \( \lambda _{2} \) for \( \tan \beta \) 
to be larger than about 2$\;$, 
where \( \lambda _{2} \) plays the role of the 
self-coupling constant of the Higgs potential in the SM\footnote{
However, 
$\tan\beta$ cannot be too large to ignore the contribution of 
the bottom quark in the case with the type-II Yukawa interaction.}. 
On the right-hand side of 
the RGE for \( \lambda _{2} \), cf. Eq.~(\ref{rge_lambda2}), there are
additional positive-definite terms \( \frac{2}{16\pi ^{2}}
\left( \lambda _{3}^{2}+
(\lambda _{3}+\lambda _{4})^{2}+\lambda _{5}^{2}+\sigma _{2}^{2}\right)  \)
as compared to the RGE for the Higgs self-coupling constant in the SM. 
These additional terms can improve vacuum stability,  
and allow lower values of \( m_{h} \). 
Therefore, one of the features of the model is to 
have a different mass range for the lightest 
CP-even Higgs boson as compared to the SM Higgs boson,
for a given cut-off scale.

\begin{figure}[t]
{\par\centering \includegraphics{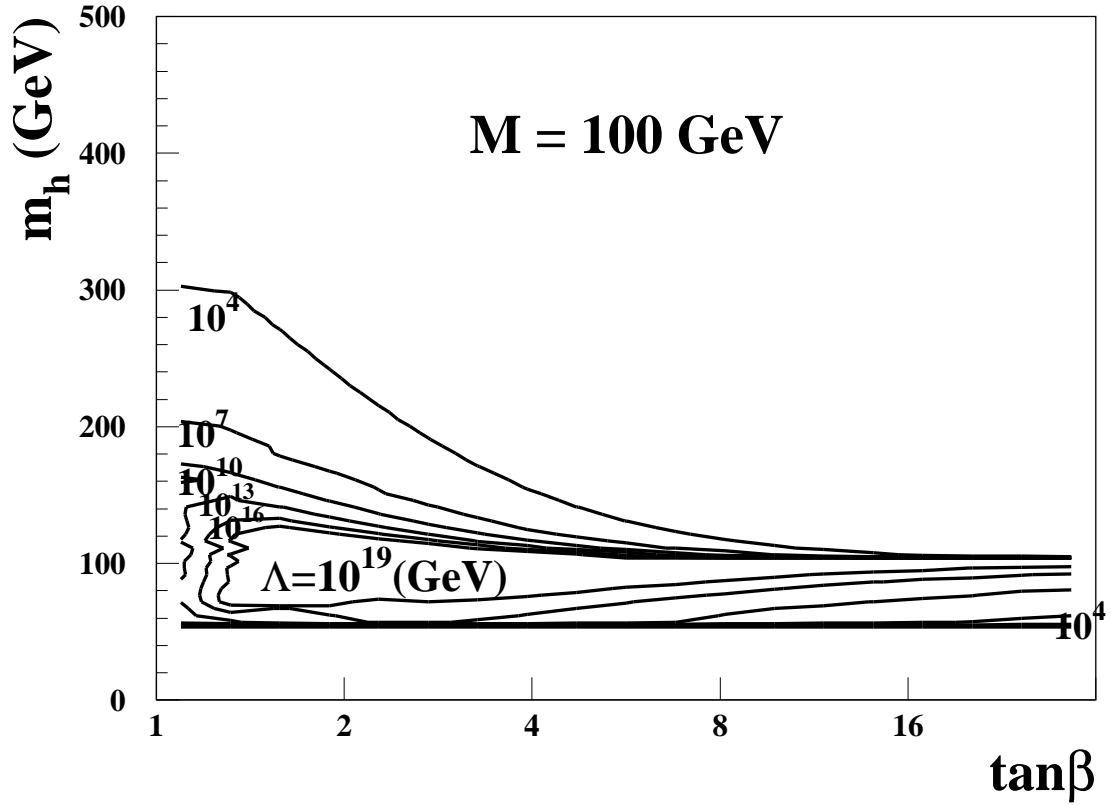} \par}
\caption{\label{fig:z100}The allowed mass range of the lightest CP-even
 Higgs boson for $M=100$ GeV.}
\end{figure}

Next, we show our result for $M$ to be around $v$.
In Fig. \ref{fig:z100}, we present the $m_h$ bound
for \( M=100 \) GeV.
In this case, the allowed range of $m_h$ is reduced 
as compared to that in the decoupling case, and
lies around $m_h \sim M$ for large $\tan \beta$. 
Notice that we have not included phenomenological constraints from the
\( b\rightarrow s\gamma  \), \( \rho  \) parameter and the direct Higgs boson
search experiment at LEP. 
As mentioned before, the mass bounds obtained from the RGE analysis
are the same for the type-I and type-II models without these
phenomenological constraints.
However, it was shown in Ref.~\cite{kko2hdm} that 
the \( b\rightarrow s\gamma  \) data 
can put a strong constraint on the allowed range
of the Higgs boson mass for \( M\lesssim 200 \)--\( 400 \) GeV 
in the type-II THDM, 
whereas there is no appreciable effect in the type-I model. 
This is because a small $M$ implies a light charged Higgs boson 
in the THDM which can induce a large
decay branching ratio for \( b\rightarrow s\gamma  \) 
in the type-II model\cite{b-sgam-2hdm}\footnote{
In addition, it has been known that the $R_b$ data also give 
strong constraints on the charged Higgs bosons 
in the type-II THDM\cite{zbb-2hdm}.}. 
We expect a similar constraint from the \( b\rightarrow s\gamma  \) 
data on the type-II Zee-model, when $M$ is small.

\begin{figure}[t]
{\par\centering \resizebox*{1\textwidth}{!}
{\includegraphics{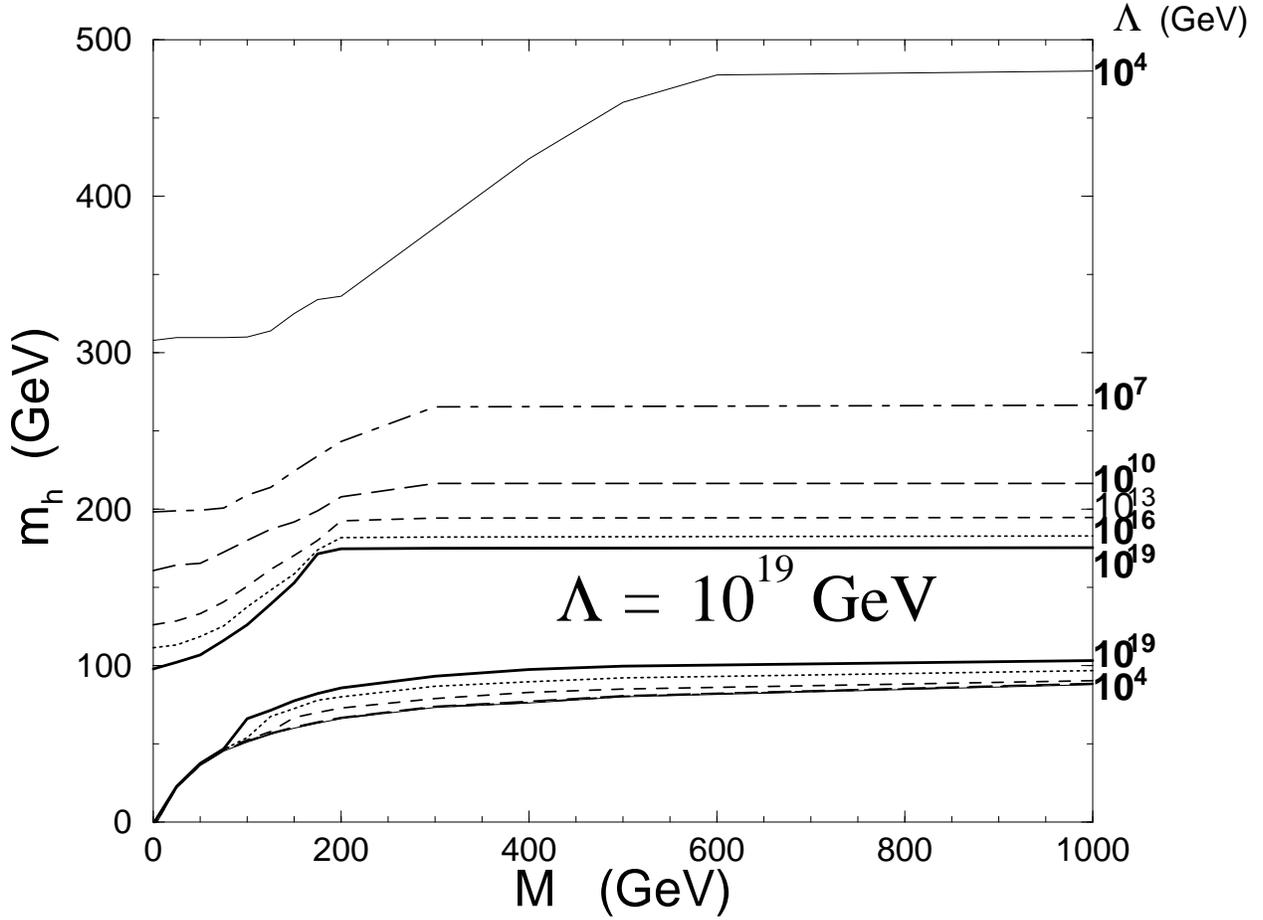}} \par}
\caption{\label{fig:zeeM}The allowed ranges of the lightest 
CP-even Higgs boson mass
as a function of \protect\( M\protect \) for various $\Lambda$ values.}
\end{figure}

In Fig.~\ref{fig:zeeM}, we show the upper and lower bounds of $m_h$ as
a function of $M$ for various values of $\Lambda$.
For given $M$, we scan the range of $\tan \beta$ for  
\( 1 \leq \tan \beta \leq 16\sqrt{2}\; (\simeq 22.6) \).
We find that the obtained $m_h$ 
bounds are almost the same as those for the THDM.
The primary reason for this is that the new coupling constants
 \( \sigma _{1} \), \( \sigma _{2} \) and \( \sigma _{3} \) do not appear
directly in the mass formula for \( m_{h} \), and therefore,
do not induce large effects
on the bounds of $m_h$.

\begin{figure}
{\par\centering \includegraphics{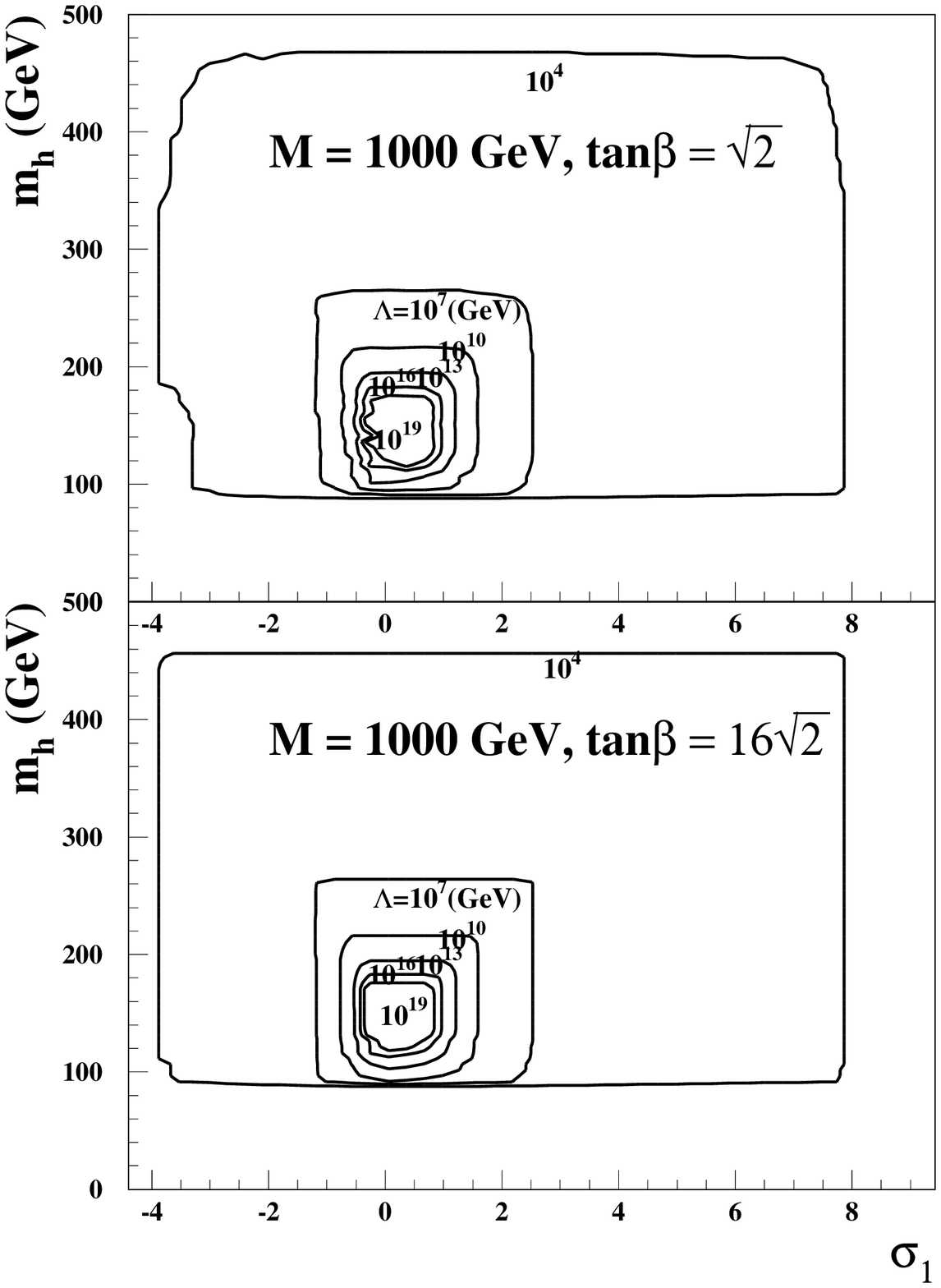} \par}
\caption{\label{fig:s1range}
The allowed range of \protect\( \sigma _{1}\protect \) and 
$m_h$ for various $\Lambda$ values.}
\end{figure}
\begin{figure}
{\par\centering \includegraphics{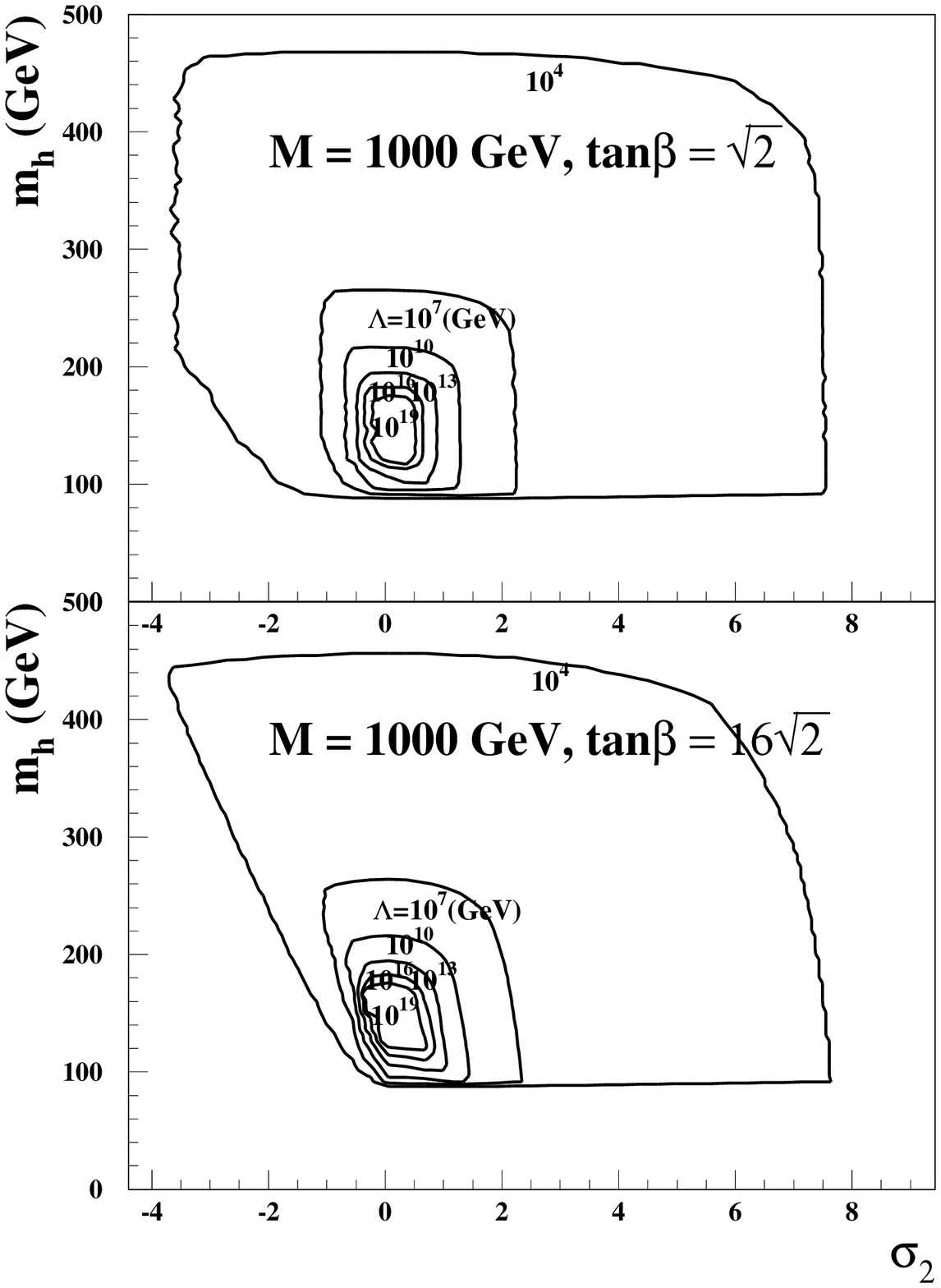} \par}
\caption{\label{fig:s2range}
The allowed range of \protect\( \sigma _{2}\protect \) and 
$m_h$ for various $\Lambda$ values.}
\end{figure}

\begin{figure}
{\par\centering \includegraphics{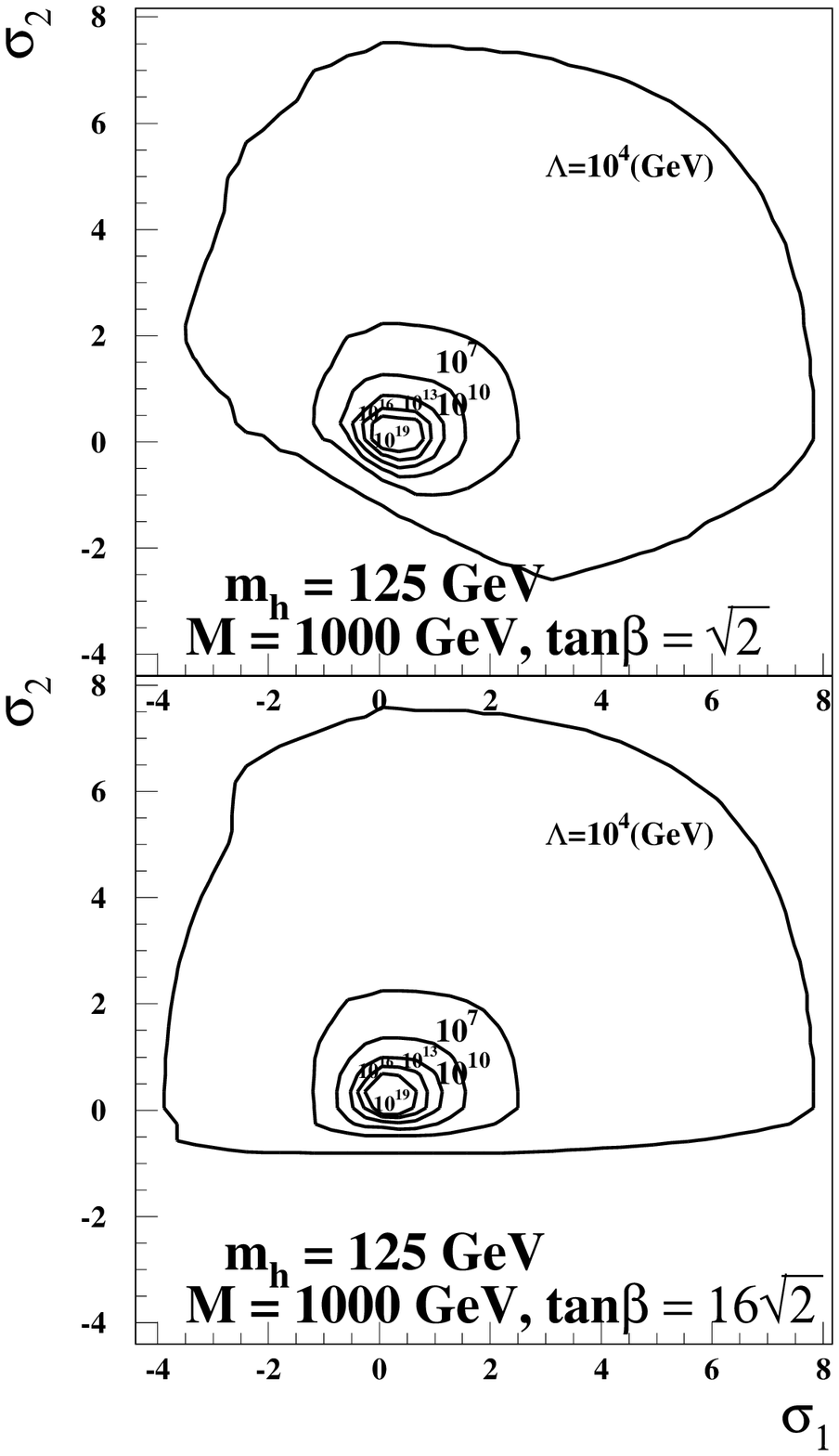} \par}
\caption{\label{fig:s12range_M1000_mh125}The allowed range 
of \protect\( \sigma _{1}\protect \)
and \protect\( \sigma _{2}\protect \) for \protect\( m_{h}=125\protect \)
GeV.}
\end{figure}

\begin{figure}
{\par\centering \includegraphics{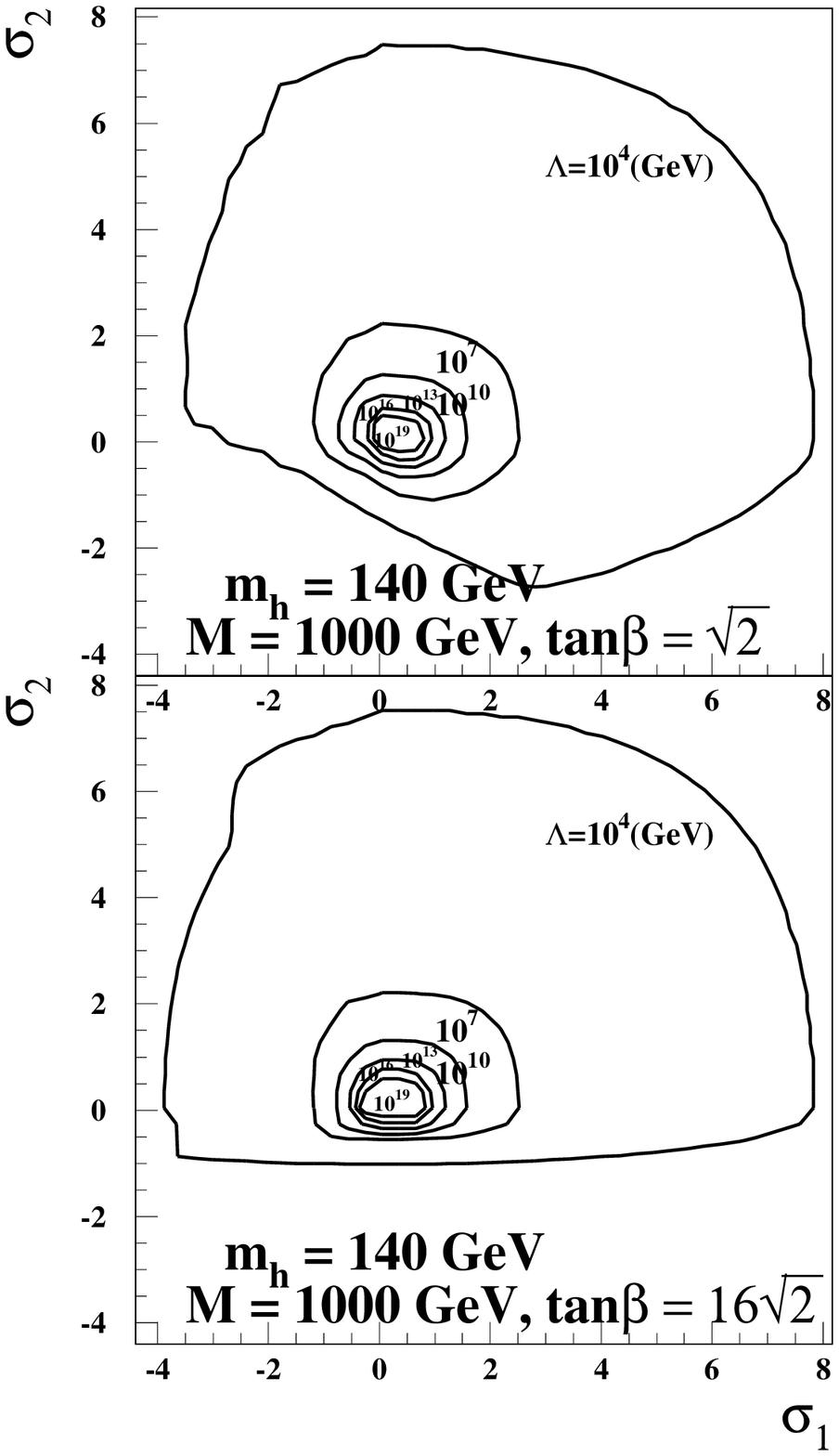} \par}
\caption{\label{fig:s12range_M1000_mh140}The allowed range 
of \protect\( \sigma _{1}\protect \)
and \protect\( \sigma _{2}\protect \) for \protect\( m_{h}=140\protect \)
GeV.}
\end{figure}

\begin{figure}
{\par\centering \includegraphics{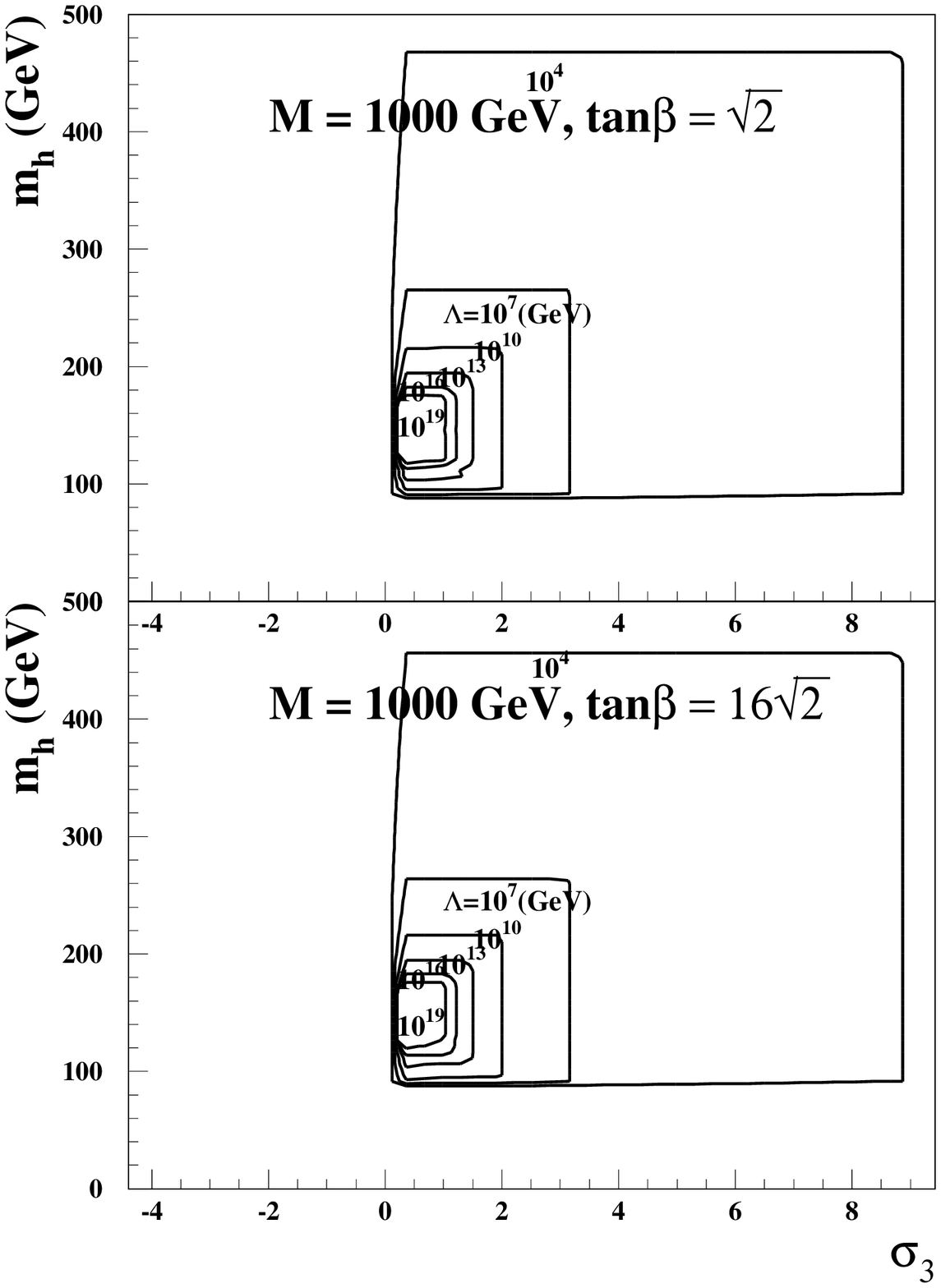} \par}
\caption{\label{fig:s3range}The allowed range of 
\protect\( \sigma _{3}\protect \)  and 
$m_h$ for various $\Lambda$ values.}
\end{figure}

We also investigate the allowed range of coupling constants \( \sigma _{1} \),
\( \sigma _{2} \) and \( \sigma _{3} \). 
For this purpose, we fix \( \sigma _{1} \)
(or \( \sigma _{2} \), \( \sigma _{3} \)) as well as \( \tan \beta  \) and
\( M \) to evaluate the upper and the lower bounds of \( m_{h} \) for each
 \( \Lambda  \) value. 
In this way, we determine the possible range of \( \sigma _{1} \) 
(or \( \sigma _{2} \), \( \sigma _{3} \))
under the condition that the theory does not break down below 
the cut-off scale
\( \Lambda  \).
In Fig.~\ref{fig:s1range}, we present the allowed 
range of \( \sigma _{1} \)
and \( m_{h} \) for different choice of \( \Lambda  \) 
in the case of \( M=1000 \)
GeV and \( \tan \beta =\sqrt{2} \) or \( 16\sqrt{2} \).
A similar figure is shown for the possible range of 
\( \sigma _{2} \) in Fig.~\ref{fig:s2range}.
We see that the maximal value of \( \sigma _{1} \) and \( \sigma _{2} \)
is around \( 0.7 \) for \( m_{h}=110-170 \) GeV if we take the cut-off scale
to be \( 10^{19} \) GeV. For smaller value of \( \Lambda  \) 
the allowed ranges
of \( \sigma _{i} \) becomes larger. For example, \( \sigma _{1} \) can exceed
\( 1 \) for \( \Lambda =10^{13} \) GeV. We have calculated for other value of
\( \tan \beta  \) and checked that these figures does not change 
greatly  between
\( \tan \beta =1.4 \) and \( 16\sqrt{2} \). 
We also present the allowed range in the \( \sigma _{1} \)
and \( \sigma _{2} \) plane for a fixed value of \( m_{h} \) 
in Figs.~\ref{fig:s12range_M1000_mh125}  and 
\ref{fig:s12range_M1000_mh140} for $m_h=125$ GeV and $m_h=140$ GeV, 
respectively.
For either value of \( m_{h} \) with \( \tan \beta =16\sqrt{2} \),
both \( \sigma _{1} \) and \( \sigma _{2} \) can be as large as 
0.5 (2)  for \( \Lambda =10^{19} \) (\( 10^{7} \))  GeV. 
The allowed range of $\sigma_3$ and $m_h$ for various values of 
$\Lambda$ is given in Fig.~\ref{fig:s3range}. 
It is shown that, $\sigma_3$ has to be larger than zero,  due to the 
vacuum stability condition. The maximal value of \( \sigma _{3} \) is about
\( 1 \) (3)  for \( \Lambda =10^{19} \) (\( 10^{7} \) )  GeV
and \( M=1000 \) GeV. 
The impact of these new coupling constants on the collider phenomenology 
is discussed in the next section.

\section{Two-photon decay width of the neutral Higgs boson}

\label{sec:h-2gamma}
In this section, we study the phenomenological consequences
of the Higgs boson mass and the Higgs-boson-coupling constants 
derived in the previous section.
The important feature of the Higgs sector of the Zee-model is that 
there are an additional weak doublet and a singlet charged Higgs boson. 
The physical states of the Higgs particles
are two CP-even Higgs bosons, one CP-odd Higgs boson and two pairs of charged
Higgs bosons. Therefore, the 
Higgs phenomenology is quite close to the ordinary 
two-Higgs-doublet model. One unique difference is the existence of 
the additional weak-singlet charged Higgs boson. 
The effect of this extra charged Higgs boson is especially
important when \( M \) is much larger than the \( Z \) boson mass,
i.e. in the decoupling regime.
 In such a case, the heavier CP-even Higgs boson, 
 the CP-odd Higgs boson as well as one
of the charged Higgs bosons have masses approximately equal to \( M \), 
and these heavy states are decoupled from low energy observables.
(Note that the condition on the applicability of the perturbation theory  
forbids too large self-couplings among the Higgs bosons. 
Hence, in the limit of large $M$, the heavy Higgs bosons decouple from 
the low energy effective theory. )
The remaining light states are the lighter CP-even Higgs boson $h$ and
the lighter charged Higgs boson $S_2$ which mainly comes from the
weak-singlet.
In the previous section, we show that, even in the decoupling case, there
can be large difference in the allowed range of $m_h$ between the 
Zee-model and the SM. Similarly, we expect that,  
even in the decoupling case, the presence of the additional weak-singlet 
charged Higgs boson can give rise to interesting Higgs phenomenology.

Since the lighter charged Higgs boson $S_2$ can couple to Higgs bosons
and leptons, it can affect the decay and the production
of the neutral Higgs bosons at colliders through radiative corrections.
In the following, we consider the decay width of \( h \to \gamma\gamma \) 
as an example.
For a SM Higgs boson, the partial decay width (or branching ratio) of
\( h\rightarrow \gamma \gamma  \) is small: 
$\sim 9.2$ keV (or $2.2 \times 10^{-3}$) for $m_h=125$ GeV, and 
$\sim 15.4$ keV (or $1.9 \times 10^{-3}$) for $m_h=140$ GeV, 
with a 175 GeV top quark. 
Nevertheless, it is an important discovery mode of the Higgs boson at
the LHC experiments for $m_h$ less than twice of the $W$-boson mass.
Needless to say that a change in the branching ratio of 
\( h \rightarrow \gamma \gamma  \) would lead to a different production 
rate of $pp \to h X \to \gamma\gamma X$.
At future $e^+e^-$ LC's, the branching ratio of $h \to \gamma\gamma$ 
can be determined via the reaction $e^+e^-\to q\overline{q} \gamma\gamma$ 
and $e^+e^-\to \nu\overline{\nu} \gamma\gamma$ with a 16-22\% 
accuracy\cite{test_e+e-lc}.   
At the photon-photon collision option of the future LC's, 
the partial decay width of \( h\rightarrow \gamma \gamma  \) 
can be precisely tested within a 2 \% accuracy\cite{test_hgaga_lc} 
by measuring the inclusive production rate of the Higgs boson $h$. 
Clearly, a change in the partial decay width of 
\( h\rightarrow \gamma \gamma  \)
will lead to a different production rate for $h$.
In the Zee-model, such a change is expected after taking
into account the loop contribution of the extra charged Higgs boson.
We find that the deviation from the SM prediction can be sizable, and 
therefore testable at the LHC and future LC's.

The partial decay width of 
\( h\rightarrow \gamma \gamma  \) is calculated at the one-loop order.
Similar to our previous discussion, we limit ourselves to the 
parameter space in which $1 \leq \tan \beta \leq 16\sqrt{2}$, and keep 
only the top quark contribution from the fermionic loop diagrams.
Including the loop contributions from the $W$ boson and the charged
Higgs bosons $S_1$ and $S_2$ together with the top quark loop
contribution, we obtain\cite{hhg}
\begin{equation}
\Gamma (h\rightarrow \gamma \gamma )=\frac{(\alpha m_{h})^{3}}
{256\pi ^{2}\sin ^{2}\theta _{W}m_{W}^{2}}
\left| \sum _{i=S_{1},S_{2},t,W}I_{i}\right| ^{2} \, ,
\end{equation}
with
\begin{eqnarray*}
I_{S_{1}} & = & R_{S_{1}}F_{0}(r_{i}) \, , \\
I_{S_{2}} & = & R_{S_{2}}F_{0}(r_{i}) \, , \\
I_{t} & = & \frac{4}{3}\left( \frac{\cos \alpha }{\sin \beta }\right) F_{1/2}(r_{i})
\, , \\
I_{W} & = & \sin (\beta -\alpha )F_{1}(r_{i}) \, ,
\end{eqnarray*}
where \( r_{i}=\frac{4m_{i}^{2}}{m_{h}^{2}} \) and \( m_{i} \) is the mass
of the internal lines in the loop diagram. 
\( R_{S_{1}} \) and \( R_{S_{2}} \) are given by
\begin{eqnarray}
R_{S_{1}} & = & \frac{v^{2}}{2}\frac{1}{m_{S_1}^{2}}\biggl 
[\cos ^{2}\chi \left\{ -\lambda _{1}\sin \alpha \sin ^{2}\beta \cos \beta +
\lambda _{2}\cos \alpha \sin \beta \cos ^{2}\beta \right. \nonumber \\
 &  & + \lambda _{3}\left( \cos \alpha \sin ^{3}\beta -
 \sin \alpha \cos ^{3}\beta \right) -
 \frac{1}{2}(\lambda _{4}+
 \lambda _{5})\cos (\alpha +\beta )\sin 2\beta \Big \}\nonumber \\
 &  & +\sin ^{2}\chi
 \left\{ -\sigma _{1}\sin \alpha \cos \beta +
 \sigma _{2}\cos \alpha \sin \beta \right\} 
+{\sqrt{2}}\sin \chi \cos \chi \frac{\mu }{v}\sin (\alpha -\beta )
 \biggr ] \, ,
\end{eqnarray}
\begin{eqnarray}
R_{S_{2}} & = & \frac{v^{2}}{2}\frac{1}{m_{S_2}^{2}}
\biggl [\sin ^{2}\chi \left\{ 
-\lambda _{1}\sin \alpha \sin ^{2}\beta \cos \beta +
\lambda _{2}\cos \alpha \sin \beta \cos ^{2}\beta \right. \nonumber \\
 &  & + \lambda _{3}\left( \cos \alpha \sin ^{3}\beta -
 \sin \alpha \cos ^{3}\beta \right) -\frac{1}{2}(\lambda _{4}+
 \lambda _{5})\cos (\alpha +\beta )\sin 2\beta \Big \}
\nonumber \\
 &  & 
+\cos ^{2}\chi \left\{ -\sigma _{1}\sin \alpha \cos \beta +
 \sigma _{2}\cos \alpha \sin \beta \right\} 
-{\sqrt{2}}\sin \chi \cos \chi \frac{\mu }{v}\sin (\alpha -\beta )
 \biggr ] \, ,
\end{eqnarray}
and
\begin{eqnarray}
F_{0}(r) & = & r\left( 1-rf(r)\right) \, , \\
F_{1/2}(r) & = & -2r\left( 1+(1-r)f(r)\right) \, , \\
F_{1}(r) & = & 2+3r+3r\left( 2-r\right) f(r) \, ,
\end{eqnarray}
with
\begin{equation}
f(r)=\left\{ \begin{array}{ll}
\left[ \sin ^{-1}\left( \sqrt{1/r}\right) \right] ^{2} & 
\mbox {if}\; r\geq 1\\
-\frac{1}{4}\left[ \ln \frac{1+\sqrt{1-r}}{1-\sqrt{1-r}}-i\pi \right] ^{2} &
 \mbox {if}\; r<1
\end{array}\right. .
\end{equation}

In the decoupling case of the model, namely \( M^{2}\gg \lambda _{i}v^{2} \),
the above formulae are greatly simplified.  
This limit corresponds to \( \alpha \rightarrow \beta -\frac{\pi }{2} \) and 
\( \chi \rightarrow 0 \), so that the light charged Higgs boson 
\( S_{2}^\pm \) is identical to the weak-singlet Higgs boson $\omega^\pm$.
Thus, we have 
\begin{equation}
R_{S_{2}}\rightarrow \frac{v^{2}}{2}\frac{1}{m_{S_2}^{2}}
\left( \sigma _{1}\cos ^{2}\beta +\sigma _{2}\sin ^{2}\beta \right) \, ,
\label{eq:decoup}
\end{equation}
and both the top-quark and the $W$ boson loop contributions reduce to 
their SM values. 
We like to stress that the weak-singlet Higgs boson 
does not directly couple to 
the quark fields in the limit of $\chi\rightarrow 0$. 
Therefore, it does not affect the decay rate of $b \ra s \gamma$ 
at the one-loop order. Similarly, being a weak singlet, it also gives 
no contribution to the $\rho$ parameter. 
Hence the low-energy constraint from either the $b \ra s \gamma$ decay
or the $\rho$ parameter on the Zee-model in the limit of $\chi \ra 0$ 
is similar to effects of that on the THDM.
\begin{figure}[t]
{\par\centering \resizebox*{1.0\textwidth}
{!}{\includegraphics{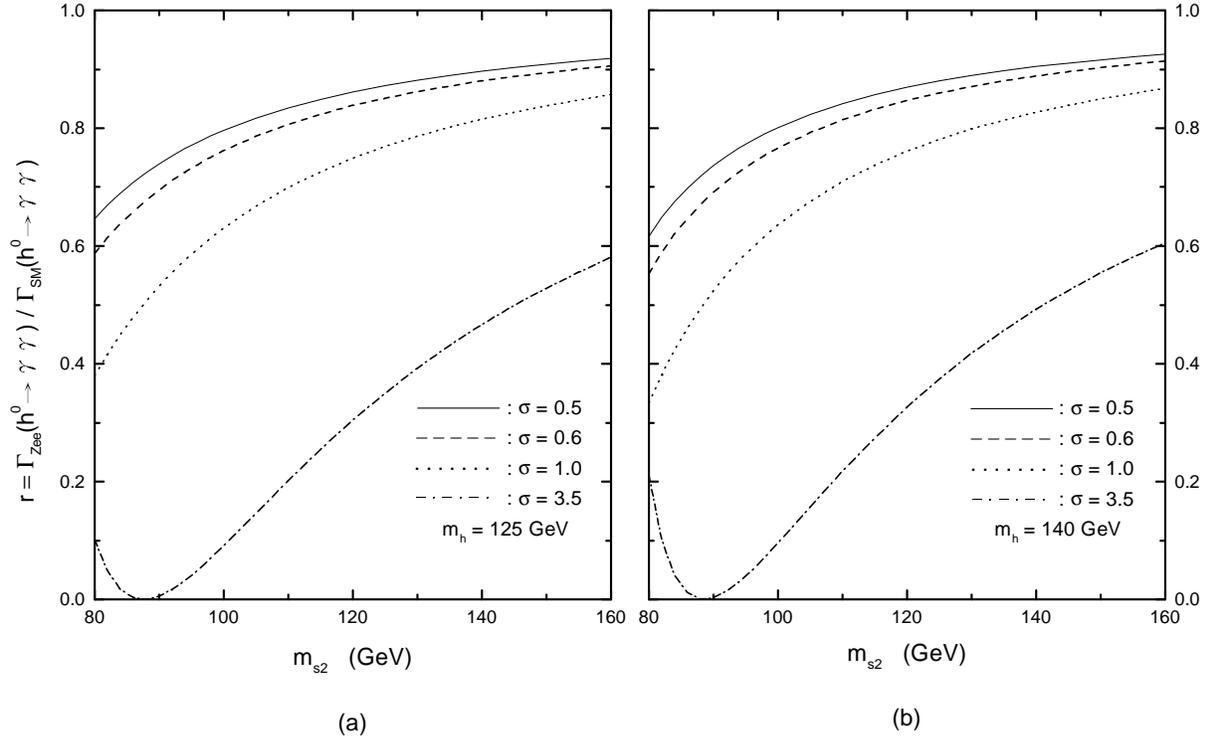}} \par}
\caption{\label{fig:ratio125_140}(a) The ratio $r$ 
as a function of the charged Higgs boson mass $m_{S_2}$ for various 
values of the coupling constants $\sigma_1=\sigma_2\equiv \sigma$ with 
$m_h=125$ GeV.
The two smaller $\sigma$'s are consistent with the cut-off scales 
$\Lambda=10^{19}$ GeV and $\Lambda=10^{16}$ GeV, respectively. 
The two larger $\sigma$'s are allowed for
$\Lambda=10^{4}$ GeV. (b) A similar plot with $m_h=140$ GeV. }
\end{figure}
\begin{figure}[t]
{\par\centering \resizebox*{1.0\textwidth}
{!}{\includegraphics{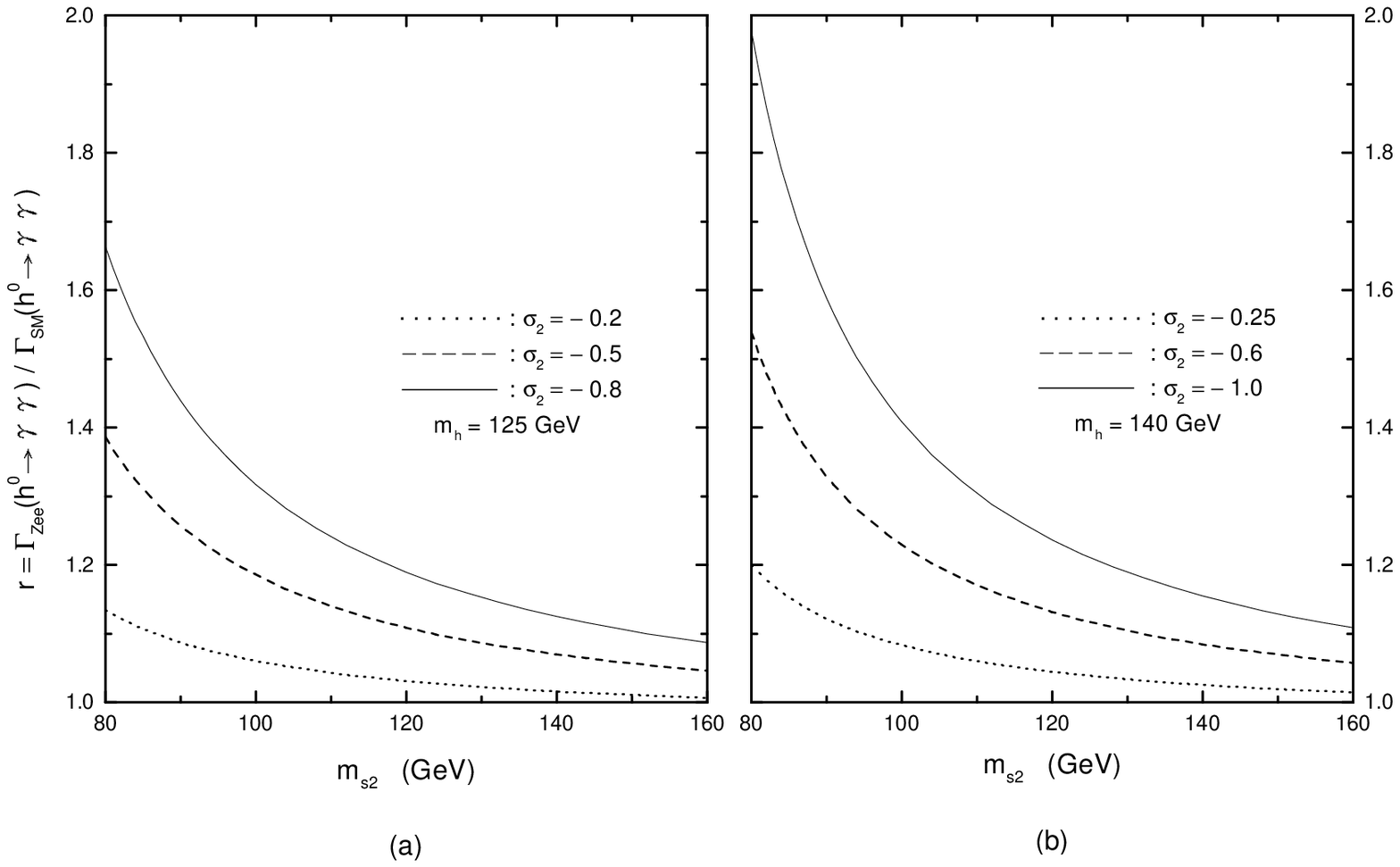}} \par}
\caption{\label{fig:ratio125_140_NegativeS2}
(a) The ratio $r$ as a function of the charged Higgs boson mass 
$m_{S_2}$ for negative values of the coupling constants 
$\sigma_2$ with $m_h=125$ GeV, $\sigma_1=0$ and $\tan\beta=16\sqrt{2}$.  
The value $\sigma_2=-0.2$, $-0.5$ or $-0.8$ is consistent 
with the cut-off scale $\Lambda=10^{19}$, $10^{7}$ or $10^4$ GeV,
respectively. 
(b) A similar plot with $m_h=140$ GeV, $\sigma_1=0$ and 
$\tan\beta=16\sqrt{2}$. 
The value $\sigma_2=-0.25$, $-0.6$ or $-1$ is consistent 
with the cut-off scale $\Lambda=10^{19}$, 
$10^{7}$ or $10^4$ GeV, respectively. 
}
\end{figure}
Let us examine at the one-loop effect of the weak-singlet charged Higgs boson 
on the decay width of $h \to \gamma\gamma$ in the decoupling limit.  
Let us recall that in Fig.~\ref{fig:s12range_M1000_mh140}, the size of 
the new couplings $\sigma_1$ and $\sigma_2$ can be as large as 2 
simultaneously, if the cut-off scale is at the order of $10^7$ GeV. 
For the Zee-model to be a valid low energy effective theory up to 
$10^{19}$ GeV, $\sigma_1$ and $\sigma_2$ cannot be much larger than 
$0.6$.  To illustrate the implications of this result,  
we show in Figs.~\ref{fig:ratio125_140} (a) 
and \ref{fig:ratio125_140} (b) the ratio ($r$) of the 
$h\rightarrow \gamma \gamma$ width predicted in the Zee-model 
to that in the SM,  
\(r \equiv \Gamma _{\rm Zee}(h\rightarrow \gamma \gamma )/
\Gamma _{\rm SM}(h\rightarrow \gamma \gamma ) \), 
as a function of the coupling constant \(\sigma_2\) and the charged Higgs 
boson mass \(m_{S_2}\). 
Here, for simplicity, we have set $\sigma_1=\sigma_2$ so that the 
$\tan \beta$ dependence drops in the decoupling case, 
cf. Eq.~(\ref{eq:decoup}).  For illustrations, we
consider two cases for the mass of the lighter CP-even Higgs boson:
$m_h=125$ GeV and $m_h=140$ GeV.
As shown in the figures, the ratio \( r \) can be around 0.8 
for \(\sigma_1=\sigma_2\equiv\sigma\approx 0.5 \) and  
\(m_{S_2}\approx 100\) GeV.
This reduction is due to the cancellation between the contribution 
from the $S_2$-boson loop and the $W$-boson loop contributions.  
To have a similar reduction rate in 
$\Gamma _{\rm Zee}(h\rightarrow \gamma \gamma )$ for a heavier $S_2$, 
the coupling constant $\sigma_2$ (and $\sigma_1$) has to be larger.
Next, as shown in Figs.~7 and 8, $\sigma_1$ and $\sigma_2$ do not 
have to take the same values in general, and they can be less than zero. 
In the case where both $\sigma_1$ and $\sigma_2$ are negative, 
the contribution of the $S_2$-loop diagram and that of the $W$-loop 
diagram have the same sign, so that $r$ can be larger than 1.  
Such an example is shown in Fig.~\ref{fig:ratio125_140_NegativeS2} (a), 
where the ratio $r$ for $m_h=125$ GeV is shown as a function of 
$m_{S_2^{}}^{}$ at various negative $\sigma_2$ values with 
$\sigma_1=0$ and $\tan\beta=16\sqrt{2}$.  
We consider the case with $\sigma_2=-0.2$, $-0.5$ or $-0.8$, 
which is consistent with the cut-off scale 
$\Lambda=10^{19}$, $10^{7}$ or $10^{4}$ GeV, respectively.  
In the case of $\Lambda=10^{19}$ GeV ($10^{4}$ GeV), the deviation from 
the SM prediction can be about +6\% (+30\%) for $m_{S_2}^{}=100$ GeV. 
In Fig.~\ref{fig:ratio125_140_NegativeS2} (b), the similar plot of 
the ratio $r$ is shown for $m_h=140$ GeV with $\sigma_1=0$ and 
$\tan\beta=16\sqrt{2}$. Each case with $\sigma_2=-0.25$, $-0.6$ or $-1$  
is consistent with $\Lambda=10^{19}$, $10^{7}$ or $10^{4}$ GeV, 
respectively. The correction is larger in the case with  
$m_h=140$ GeV than in the case with $m_h=125$ GeV for a given $\Lambda$.  
The deviation from the SM prediction can amount to about +8\% (+40\%) 
for $\Lambda=10^{19}$ GeV ($10^{4}$ GeV) when $m_{S_2^{}}^{}=100$ GeV. 
Larger positive corrections are obtained for smaller $m_{S_2^{}}^{}$ values.  
Such a deviation from the SM prediction can be tested at the LHC, 
the $e^+e^-$ LC and the $\gamma\gamma$ option of LC. 

\begin{figure}[t]
{\par\centering \resizebox*{0.6\textwidth}
{!}{\includegraphics{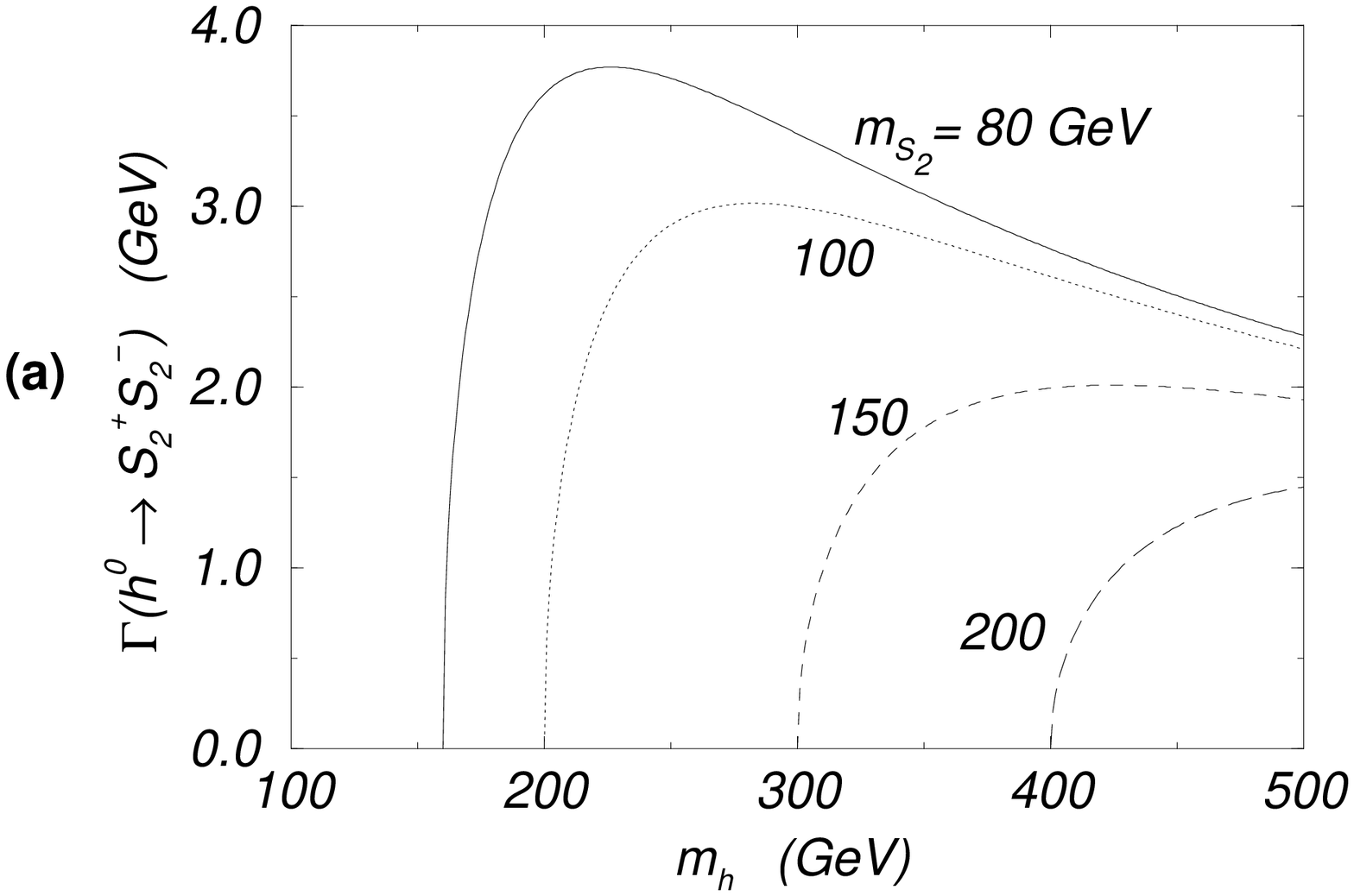}} \par}
{\par\centering \resizebox*{0.6\textwidth}
{!}{\includegraphics{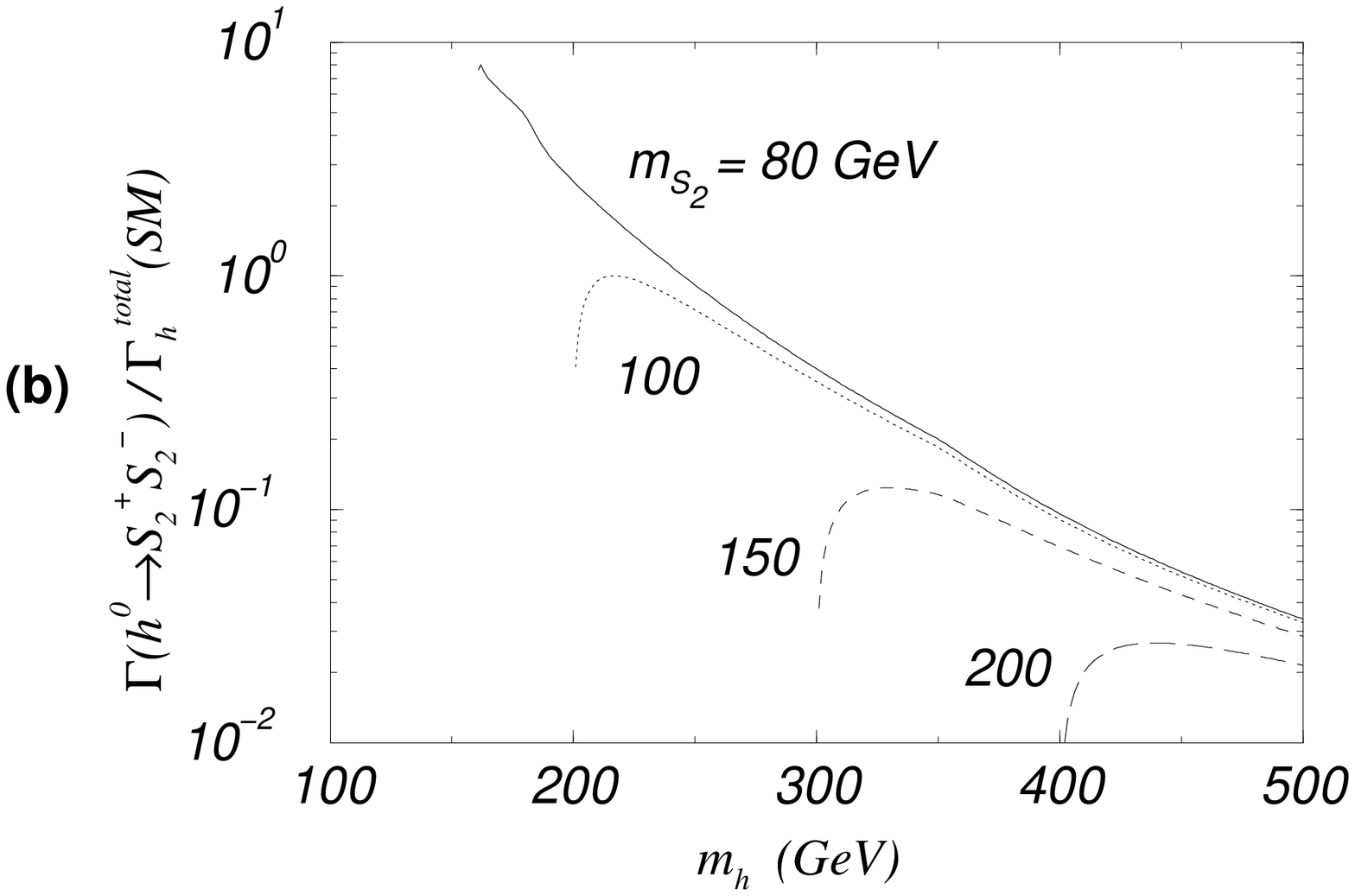}} \par}
\caption{\label{fig:hsswidth} 
(a) The partial decay width $\Gamma(h \ra S^+_2 S^-_2)$ 
for $m_{S_2}=80,100,150,200$ GeV with $\sigma_1=\sigma_2=1$
for the allowed range of $m_h$ from 100 GeV to 500 GeV. 
(b) The ratio of $\Gamma(h \ra S^+_2 S^-_2)$ with  
    the total decay width of the SM Higgs boson for each values 
    of $m_{S_2}^{}$. }
\end{figure}
Before concluding this section, we remark that if $m_h$ is larger than 
$2 m_{S_2}$ such that the decay mode $ h \ra S^+_2 S^-_2$ is open, 
the total decay width of $h$ can be largely modified from the SM
prediction for large $\sigma_{1,2}$. In terms of $R_{S_2}$, the partial 
decay width of $h \ra S^+_2 S^-_2$ is given by 
\begin{equation}
\Gamma(h \ra S^+_2 S^-_2) = {c^2 v^2 \over 16 \pi m_h} 
\sqrt{1-{4 m^2_{S_2} \over m^2_h}} \, ,
\label{eq:hs2s2}
\end{equation}
where $c^2=(2 m^2_{S_2} R_{S_2}/v^2)^2$.
In Fig.~\ref{fig:hsswidth} (a), we show the partial decay width 
$\Gamma(h \ra S^+_2 S^-_2)$ for $m_{S_2}^{}=80,100,150,200$ GeV 
with $\sigma_1=\sigma_2=1$, cf. Eq.~(\ref{eq:decoup}), 
for the allowed range of $m_h$ from 100 GeV
to 500 GeV. In Fig.~\ref{fig:hsswidth} (b), 
the ratio of $\Gamma(h \ra S^+_2 S^-_2)$ to the total width 
of the SM Higgs boson ($\Gamma_{h}^{\rm total}(SM)$) is shown as a 
function of $m_h$ for each value of $m_{S_2}^{}$. This is to illustrate 
the possible size of the difference between the total width of the 
lightest CP-even Higgs boson $h$ in the Zee-model and that of the 
SM Higgs boson\footnote{
In doing this analysis, we have in mind a low cut-off
scale $\Lambda=10^4$ GeV, which allows a wide range of values
for $\sigma$'s, $m_{S_2}$ and $m_h$. 
}.
Clearly, the impact of the $S_2^+S_2^-$ decay channel is especially large 
in the small $m_h$ region.  
We note that $\Gamma_{h}^{\rm total}(SM)$ can be determined 
to the accuracy of 10-20\% at the LHC and the LC if 
$m_h < 2 m_Z^{}$, and to that of a couple of per cents 
if $m_h > 2 m_Z^{}$\cite{sopczak}. ($m_Z^{}$ is the mass of the $Z$-boson.)
Hence, measuring the total width of the lightest neutral 
Higgs boson can provide a further test of the Zee-model 
for $m_h^{} > 2 m_{S_2}^{}$.
The change in the total width  also modifies the decay branching
ratio of $h \ra Z Z$, hence yields a different rate 
of $h \ra Z Z \ra \mu^+ \mu^- \mu^+ \mu^-$ for a given $m_h$.
(In the SM, the branching ratio of $h \ra Z Z$ is about 1/3 for 
$m_h > 200$ GeV.)
Needless to say that for $m_h > 2 m_{S_2}$, the production mode of
$h \ra S^+_2 S^-_2 \ra \ell^+ {\ell'}^- \etv$ 
is also useful to test the Zee-model. 
Further discussion on this possibility will be given in Sec.~VI.

\section{Phenomenology of charged-Higgs bosons}

\label{sec:charged}
In the Zee-model, two kinds of charged Higgs bosons appear.   
If there is no mixing between them ($\chi=0$), the mass eigenstates 
$S_1^\pm$ and $S_2^\pm$ correspond to the THDM-like charged Higgs field 
and the singlet Higgs field $\omega^\pm$, respectively.  
The case with $\chi=0$ occurs in the limit of 
$M^2 \gg v^2, \mu^2$ and $m_0^2$; i.e. in the decoupling limit.   
The  detection of $S_2^\pm$ can be a clear indication of the Zee-model.
As to be shown later, its phenomenology is found to be drastically different  
from that of the THDM-like charged Higgs bosons $S_1^\pm$\cite{zee-prl}.  
Here, we discuss how the effects of this extra charged boson can be 
explored experimentally.
We first consider the case with $\chi=0$, and then extend the discussion 
to the case with a non-zero $\chi$.

The $S_2^-$ boson decays into a lepton pair $e_i^- \overline{\nu}_{e_j}^c$ 
with the coupling constant $f_{ij}$. The partial decay rate, 
$\Gamma^{S_2}_{ij}=\Gamma(S_2^- \to e_i^- \overline{\nu}_{e_j}^c)$,  
is calculated as 
\begin{eqnarray}
 \Gamma^{S_2}_{ij} = \frac{m_{S_2}^{}}{4\pi} \, f_{ij}\,^2 
             \left( 1 - \frac{m_{e_i}^2}{m_{S_2}^2} \right)^2,   
\end{eqnarray}
and the total decay width of $S_2^-$ is given by 
\begin{eqnarray}\label{s2width_0}
 \Gamma^{S_2}_{\rm total} = 
\sum_{i,j=1}^3 \Gamma^{S_2}_{ij} .
\end{eqnarray}
By taking into account the hierarchy pattern of $f_{ij}$, cf. 
Eqs.~(\ref{eq21}) and (\ref{eq22}),   
and by assuming $m_{S_2}^{}=100$ GeV and $|f_{12}|=3\times 10^{-4}$, 
the total decay width and the life time ($\tau$) 
is estimated to be\footnote{
The size of the decay width depends on the value of $f_{12}$. 
If we take $m_{S_1} > 500$ GeV or $\mu < 100$ GeV, $f_{12}$ can become 
one order of magnitude larger 
than $3 \times 10^{-4}$, while still being consistent with the 
phenomenological bounds discussed in Sec.~II.}
\begin{eqnarray}
\Gamma^{S_2}_{\rm total} 
&\sim& \Gamma^{S_2}_{12} + \Gamma^{S_2}_{21} 
\sim 1.6 \; {\rm keV}, \;\;  \\ 
\tau &\sim& 1/\Gamma_{\rm total}^{S_2} \sim 10^{-18} {\rm sec}. 
\end{eqnarray}
This implies that $S_2$ decays after traveling a distance of 
$\sim 10^{-10}$ m, which is significantly shorter than 
the typical detector scale. 
Therefore, $S_2^\pm$ decays promptly after its production, and can be 
detected at collider experiments.

The main production channel at the LEP-II experiment may be the pair 
production process 
$e^+e^- \to S^+_2 S^-_2$, similar to the production of 
the THDM-like charged Higgs boson $S_1^+$.   
The matrix-element squares for the $S_i^+S_i^-$ production 
($i=1,2$) are given by   
\begin{eqnarray}
&& 
\left| {\cal M}
(e^-_{L(R)} e^+_{R(L)} \to S_i^+S_i^-) \right|^2 =          
\left\{ \frac{Q_e^{} e^2}{s}    - 
                  \frac{1}{c_W^2}                    
                 (I_{S_i}^3 - s_W^2 Q_{S_i}^{}) 
                  \frac{(I_e^3 - s_W^2 Q_e) g^2}{s - m_Z^2} \right\}^2 
s^2 \beta_{S_i}^2 \sin^2 \Theta,  
\end{eqnarray}
where $Q_e=-1$ and $I_e^3 = - \frac{1}{2}$ ($0$) for the 
incoming electron $e_L^-$ ($e_R^-$); 
$Q_{S_i}^{}=-1$ and $I_{S_i}^3 = - \frac{1}{2}$ ($0$) 
for $i=1$ ($2$); $\beta_{S_i^{}}=\sqrt{ 1 - 4 m_{S_i}^2/s}$, 
$s_W=\sin \theta_W$, $c_W = \cos \theta_W$, and 
$\Theta$ is the scattering angle of $S_i^-$ in the $e^+e^-$ 
center-of-mass (CM) frame whose energy is $\sqrt{s}$. 
For the other electron-positron helicity configuration   
($e^-_Le^+_L$ and $e^-_Re^+_R$), the cross sections are zero. 
Thus the total cross section for the $S_2^+S_2^-$ pair production is given by 
\begin{eqnarray}
 \sigma(e^+e^- \to S^+_2 S^-_2) = 
     \frac{1}{96\pi} \,e^4  \beta^3_{S_2}  s   
     \left[ \left( \frac{1}{s} 
             + \frac{s_W^2}{c_W^2} \frac{1}{s - m_Z^2} \right)^2  
     +       \left\{ \frac{1}{s} 
              - \left( \frac{1}{2} - s_W^2 \right) 
  \frac{1}{c_W^2} \frac{1}{s - m_Z^2} \right\}^2  \right]. 
\end{eqnarray}
Hence, the production rates of $S_1^-$ and $S_2^-$ are different. 
We note that the ratio of cross sections for $S^+_1 S^-_1$ and 
$S^+_2 S^-_2$ production,
$\sigma(e^+e^- \to S^+_2 S^-_2)/\sigma(e^+e^- \to S^+_1 S^-_1)$, 
is $0.8$ at $\sqrt{s}=210$ GeV 
assuming that the masses of $S_1^\pm$ and $S_2^\pm$ are the same. 
This ratio is independent of the masses of $S_1^{}$ and $S_2^{}$ 
for a fixed CM energy.  
(Only the difference between  $S_1^+S_1^-Z$ and $S_2^+S_2^-Z$ 
coupling constants determines this ratio. )

\begin{figure}[t]
{\par\centering \resizebox*{0.6\textwidth}
{!}{\includegraphics{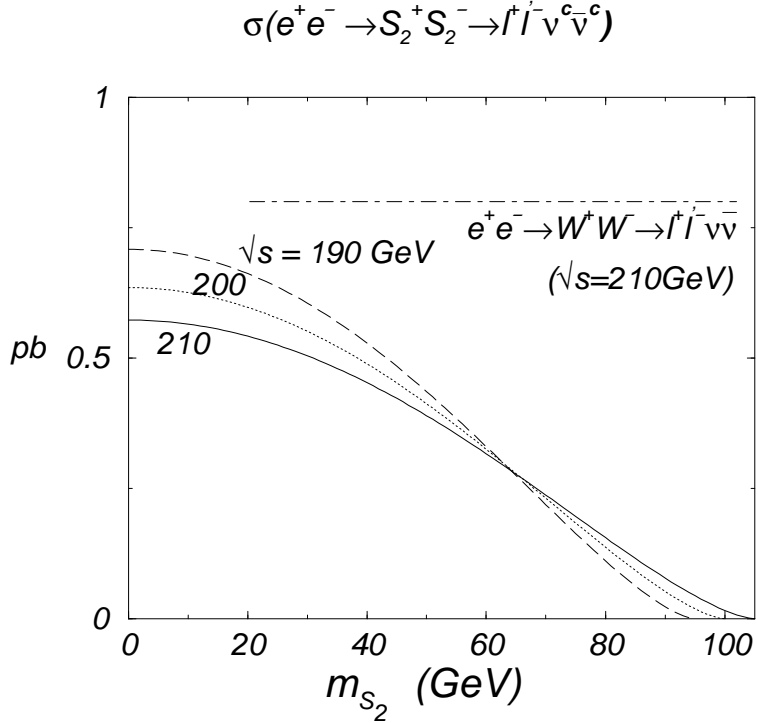}} \par}
\caption{\label{fig:zee_pair}
The cross section of the leptonic decay process 
$e^+e^- \to S_2^+S_2^- \to \ell^+\ell'^- 
\etv$ (where $\ell$ and $\ell'= e$ or $\mu$) 
at $\sqrt{s}=190$, 200, 210 GeV. 
The process $e^+e^- \to W^+W^- 
\to \ell^+ \ell'^- \etv$ 
at $\sqrt{s}=210$GeV is shown for comparison. }
\end{figure}

The lower mass bound of the THDM-like charged boson 
$S^\pm_1$ can be obtained by studying its $\tau \nu$ and $c s$ decay 
modes, completely in the same way as the charged Higgs boson search in the 
minimal supersymmetric standard model (MSSM)\cite{charged-search}. 
Similar experimental constraints may be obtained for the extra charged 
bosons $S_2^\pm$. The situation, however, turns out to be fairly different 
from the $S_1^\pm$ case.   
First of all, decays of $S_2^\pm$ are all leptonic. 
Secondly, the branching ratios of various $S_2^\pm$ decay modes  
are estimated as 
\begin{eqnarray}
  B(S_2^- \to e^- \etv) &\sim& 0.5, \\
  B(S_2^- \to \mu^- \etv) &\sim& 0.5, \\
  B(S_2^- \to \tau^- \etv) &\sim& 
  {\cal O}\left( \frac{m_\mu^4}{m_\tau^4} \right) \sim 10^{-5},  
\end{eqnarray}
where we have used the relations given in Eqs. (\ref{eq21}) and (\ref{eq22}).
Clearly, the branching ratio into the $\tau^- \etv$ mode is very small, 
so that it is not useful for detecting $S_2^\pm$ at all.  
This is different from the case of detecting the ordinary THDM-like 
charged Higgs boson, which preferentially decays into heavy fermion pairs 
(e.g. $\tau \nu$ and $cs$).  
Instead of studying the $\tau^\pm {\nu}^c$ mode, the $e^\pm\, \nu^c$ and 
$\mu^\pm\,\nu^c$ modes can provide a strong constraint on the mass of 
$S_2^\pm$. In fact, the branching ratio of $S_2^- \to e^- \etv$ or 
$\mu^- \etv$ is almost 100 \%, so that we have $\sigma(e^+e^- \to S_2^+S_2^- 
 \to \ell^+ \ell'^- \etv) \sim \sigma(e^+e^- \to S_2^+S_2^-)$, where 
$\ell^-$ and $\ell'^-$ represent $e^-$ or $\mu^-$ (not $\tau^-$). 
Let us compare this with the cross section 
$\sigma(e^+e^- \to W^+W^- \to \ell^+ \ell'^- \etv ) = 
 \sigma(e^+e^- \to W^+W^-) \cdot B(W^- \to \ell^- \etv)^2$, 
where 
$B(W^- \to \ell^- \etv)
 = B(W^- \to e^- \etv) + B(W^- \to \mu^- \etv) 
\sim 21 \%$. 
As seen in Fig.~\ref{fig:zee_pair},  the cross section 
$\sigma(e^+e^- \to S_2^+S_2^- \to \ell^+ \ell'^- \etv)$ 
is comparable with $\sigma(e^+e^- \to W^+W^- 
\to \ell^+ \ell'^- \etv)$. 
Therefore, by examining the LEP-II data for $\ell^+ \ell'^- \etv$ 
(where $\ell^+ \ell'^- = e^+e^-$, $e^\pm\mu^\mp$ or $\mu^+\mu^-$,  
in contrast to $\tau^+\tau^-$ for the $S_1^\pm$ case), 
the experimental lower bound on the mass of $S_2^\pm$ can be determined. 
Such a bound can be induced from the smuon search 
results at the LEP experiments\cite{slepton,slepton2} 
in the case that neutralinos are assumed to be massless. 
From the $\mu^+\mu^-\etv$ data accumulated up to $\sqrt{s}=202$ 
GeV\cite{slepton2}, 
we find that the lower mass bound of $S_2^\pm$ is likely to be 
80-85 GeV for the $\chi=0$ cases.   
[ We note that the right-handed smuon ($\tilde{\mu}_R^\pm$) in the MSSM 
carries the same $SU(2) \times U(1)$ quantum number as 
the weak-singlet charged Higgs boson ($S_2^\pm$ for $\chi^\pm\sim 0$). ]  
 
\begin{figure}[hpth]
{\par\centering \resizebox*{0.6\textwidth}
{!}{\includegraphics{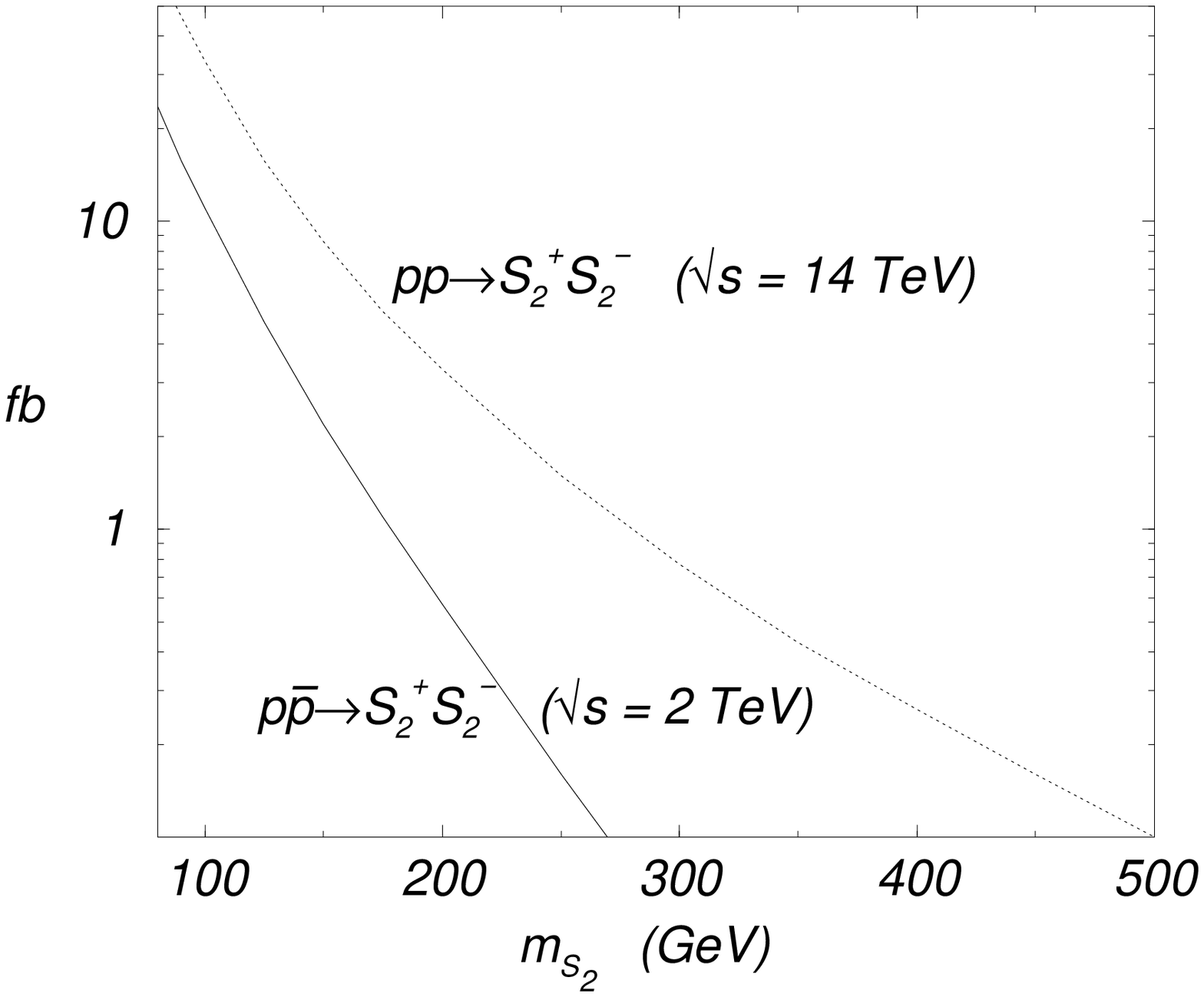}} \par}
\caption{\label{fig:zee_hadron}
The total cross sections of 
$p \overline{p} \to S^+_2S^-_2$ at $\sqrt{s}=2$ TeV (solid curve) 
and $p p \to S^+_2S^-_2$ at $\sqrt{s}=14$ TeV (dotted curve) 
as a function of $m_{S_2}^{}$.}
{\par\centering \resizebox*{0.6\textwidth}
{!}{\includegraphics{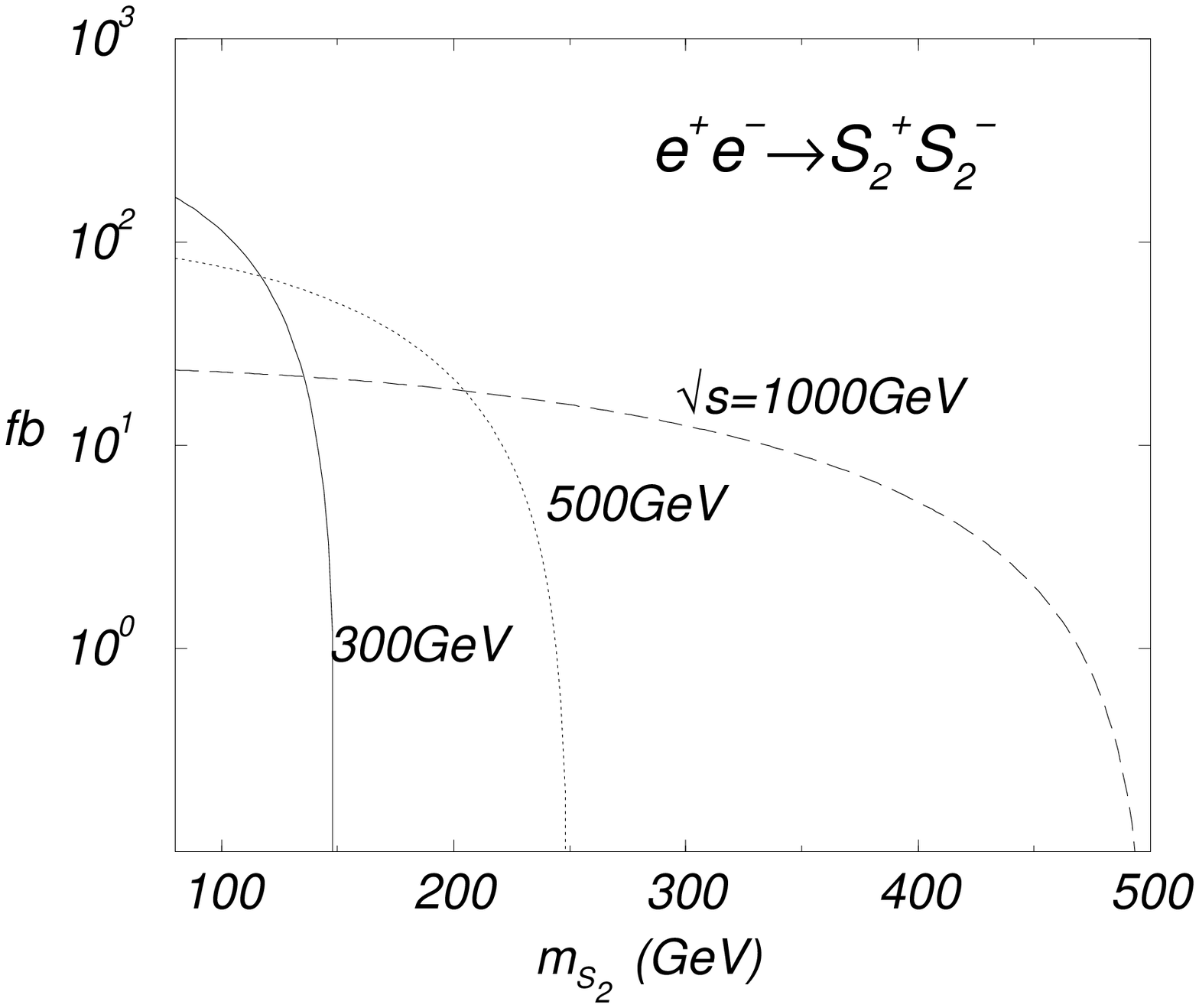}} \par}
\caption{\label{fig:zee_LC}
The total cross section of $e^+e^- \to S^+_2S^-_2$ 
as a function of $m_{S_2}^{}$ at $\sqrt{s} = 300$, 500 and 1000 GeV.}
\end{figure}

We next comment on $S_2^\pm$-production processes 
at hadron colliders and future LC's. 
At hadron colliders, the dominant production mode 
is the pair production through the Drell-Yan-type process.    
The cross sections for $p \overline{p} \to S_2^+S_2^-$ 
at the Tevatron Run-II energy ($\sqrt{s}=2$ TeV) and 
$p p \to S_2^+S_2^-$ at the LHC energy ($\sqrt{s} = 14$ TeV) 
are shown as a function of $m_{S_2}$ in Fig.~\ref{fig:zee_hadron} 
for $\chi = 0$.   
At future LC's, the $S_2^\pm$ boson may be discovered through the 
above-discussed pair-production process from the electron-positron 
annihilation if $\sqrt{s}/2 > m_{S_2}$.  
In Fig.~\ref{fig:zee_LC}, we show the total cross section of   
$e^+e^- \to S_2^+S_2^-$ for $\chi=0$ as a function of $m_{S_2}$ for 
$\sqrt{s}=300$, $500$, and $1000$ GeV.

\begin{figure}[t]
{\par\centering \resizebox*{0.6\textwidth}
{!}{\includegraphics{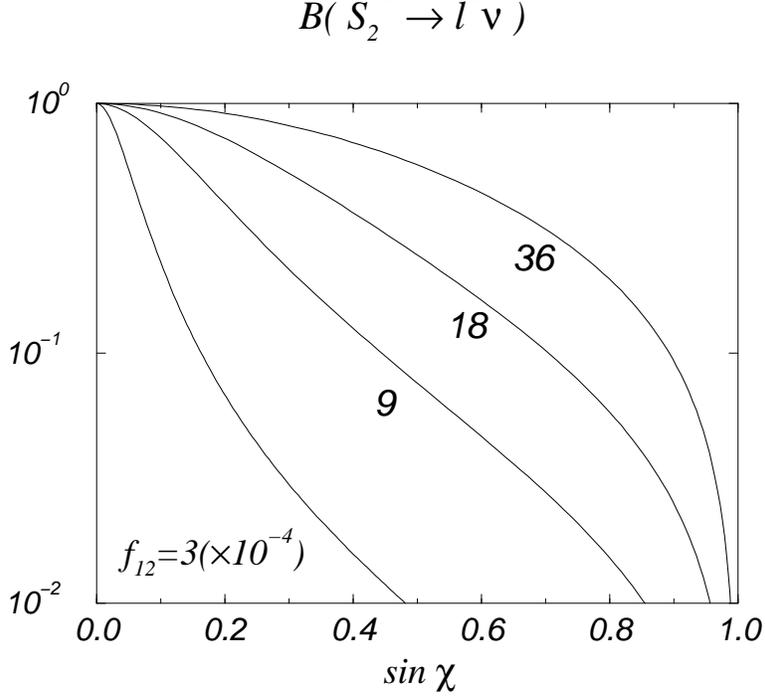}} \par}
\caption{\label{fig:nonmix}
\noindent
The decay branching ratio 
of $S_2^- \to \ell^- \etv$ (where $\ell^- = e^-$ or $\mu^-$)
as a function of the mixing angle $\chi$ for 
$m_{S_2^{}} = 100$ GeV, $\tan\beta=1$ 
and various values of the coupling constant $f_{12}$.    }
\end{figure}

Finally, we like to discuss the case with a non-zero $\chi$, in which 
$S_2^-$ is a mixture of the singlet charged Higgs boson 
state ($\omega^-$) and the doublet charged Higgs boson state ($H^-$). 
Let us see how the above discussion is changed 
in this case. The doublet charged Higgs bosons with the mass 
of $100$ GeV mainly decay into the 
$\tau^- \nu$ and $\overline{c} s$ channels. 
Thus, the branching ratio of the decay process 
$S_2^- \to \ell^- \etv$, where $\ell^-$ represents $e^-$ and $\mu^-$, 
is expressed in a non-zero $\chi$ case as 
\begin{eqnarray}
  B(S_2^- \to \ell^- \etv) 
  = \frac{\cos^2 \chi \,\, \Gamma_{\rm total}^{S_2}|_{\chi=0}}
         {\sin^2 \chi \,\, \Gamma_{\rm total}^{S_1}|_{\chi=0}
        + \cos^2 \chi \,\, \Gamma_{\rm total}^{S_2}|_{\chi=0}},   
\end{eqnarray}
where $\Gamma_{\rm total}^{S_i}|_{\chi=0}$ ($i = 1,2$) is  
the total width of $S_i^-$ at $\chi=0$ with the same mass as the 
decaying $S_2^-$ on the left-hand side of the above equation. The formula of 
$\Gamma_{\rm total}^{S_2}|_{\chi=0}$ is given in Eq.~(\ref{s2width_0}), 
while $\Gamma_{\rm total}^{S_1}|_{\chi=0}$, which is the same as 
the total decay width of the charged Higgs boson in the THDM is 
given by 
\begin{eqnarray}
  \Gamma_{\rm total}^{S_1}|_{\chi=0} = 
   \sum_{\overline{f}f'} \Gamma(S_1^- \to \overline{f}f'),   
\end{eqnarray}
where $\overline{f}f'$ are fermion pairs which are kinematically allowed. 
In the type-II Yukawa couplings, we have 
\begin{eqnarray}
\Gamma(S_1^- \to \tau^- \nu) &=& 
   \frac{m_{S_1^{}}}{8\pi v^2} (m_\tau^2 \tan^2 \beta) 
   \left(1 - \frac{m_\tau^2}{m_{S_1^{}}^2}\right)^2,\\  
\Gamma(S_1^- \to \overline{c} s) &\simeq& 
   \frac{3 m_{S_1^{}}}{8\pi v^2} (m_s^2 \tan^2 \beta + m_c^2 \cot^2 \beta) 
   \left(1 - \frac{m_c^2}{m_{S_1^{}}^2}\right)^2,\\  
\Gamma(S_1^- \to \overline{t} b) &\simeq& 
   \frac{3 m_{S_1^{}}}{8\pi v^2} (m_b^2 \tan^2 \beta + m_t^2 \cot^2 \beta) 
   \left(1 - \frac{m_t^2}{m_{S_1^{}}^2}\right)^2. 
\end{eqnarray} 
In the THDM, the total decay width of the charged Higgs boson ($H^-$) 
for $m_{H^-} = 100$ GeV is about 470 keV. Hence, if the mixing  
angle $\chi$ is not so small, the decay pattern of $S_2^-$ is dominated 
by that of the THDM charged Higgs boson $H^-$.   
In Fig.~\ref{fig:nonmix},  
we plot the branching ratio 
$B(S_2^- \to \ell^- \etv)$ as a function of $\sin\chi$ 
at $m_{S_2}^{}=100$ GeV for several values of $f_{12}$.   
We only show the case with $\tan\beta=1$, where the result is independent 
of the type of the Yukawa interaction. 
The coupling constant $f_{12}$ is taken to be 
$3, 9, 18$ and $36$ ($\times 10^{-4}$), which satisfy the phenomenological 
constraints given in Sec.~II. As expected, 
the branching ratio decreases as $\chi$ increases. 
When $f_{12}=36 \times 10^{-4}$, $B(S_2^- \to \ell^- \etv)$ 
is smaller than 10\% for $\sin\chi > 0.89$.
For the smaller $f_{12}$ values, the branching ratio reduces more quickly. 
The branching ratio is not sensitive to $m_{S_2^{}}^{}$  
unless the mass exceeds the threshold of the decay into 
a $\overline{t} b$ or $h^0 W^-$ pair. 
Above the threshold of the $\overline{t} b$ pair production, 
the decay rate of $S_2^- \to \overline{t} b$ is large due to the 
large mass of the top quarks,  
so that $B(S_2^- \to \ell^- \etv)$ is substantial only for very  
small values of $\chi$. 
Finally, while the decay branching ratio can change drastically depending on 
the mixing angle $\chi$, the production cross section for 
$e^+e^- \to S_2^+S_2^-$ remains unchanged. 
In conclusion, the process $e^+e^- \to S_2^+S_2^- \to \ell^+\ell'^- \etv$ 
can also be useful for testing the Zee-model in the non-zero 
$\chi$ case, provided $\sin\chi$ is not too large.

\section{Discussion and Conclusion}

\label{sec:conclusion}
In this paper, the Higgs sector of the Zee-model has been 
investigated, in which neutrino masses are generated radiatively. 
This model contains extra weak-doublet Higgs field and singlet 
charged Higgs field. 

We have studied indirect effects of these extra Higgs bosons 
on the theoretical mass bounds of the lightest CP-even Higgs boson, 
which are obtained from the requirement that the running coupling 
constants neither blow up to a very large value nor fall down to 
a negative value, up to a high-energy cut-off scale \( \Lambda  \).  
For \( \Lambda =10^{19} \) GeV, the upper bound of \( m_{h} \) is 
found to be about \( 175 \) GeV, 
which is almost the same value as the SM prediction. 
In the decoupling regime ($M \gg m_Z^{}$), the lower bound is found  
 to be about \( 100 \) GeV for \( \Lambda =10^{19} \) GeV, which 
is much smaller than the lower bound in the SM,  and is almost the same 
as that in the THDM.  
For smaller \( \Lambda \) values, the bounds are more 
relaxed, similar to that of the SM.      
We have also investigated the allowed range of the coupling constants 
relevant to the weak-singlet Higgs field.

The most striking feature of the Zee-model Higgs sector is the existence 
of the weak-singlet charged Higgs boson. 
We have examined the possible impact of the singlet charged-Higgs 
boson on the neutral Higgs boson search through radiative corrections. 
We found that its one-loop contributions to the 
\( h\rightarrow \gamma \gamma  \) width can be sizable. 
In the allowed range of the coupling constants the deviation 
from the SM prediction for this decay width can be 
about $-20$\% or near $+10$\% for $m_{S_2}^{}=100$ GeV 
and $\Lambda=10^{19}$ GeV, depending on the sign of the coupling 
constants $\sigma_i$.
The magnitude of the deviation is larger for 
lower $\Lambda$ values or for smaller $m_{S_2}$ values. 
For example, a positive deviation over $30$-$40$\% is possible    
for $m_h^{} = 125$-$140$ GeV, $m_{S_2}^{} = 100$ GeV, and
$\Lambda = 10^4$ GeV.  

In the decoupling limit (i.e. when $M^2 \gg v^2$, where 
$\alpha \to \beta - \pi/2$ and $\chi \to 0$), we expect that the 
production cross sections for $gg \to h$, $e^+e^- \to \nu\overline{\nu} h$ 
and $e^+e^- \to Z^0 h$ in the Zee-model are the same as those in the SM. 
However, a sizable change in the decay branching ratio of 
$h \to \gamma\gamma$ can alter the production rate of 
$pp \to h X \to \gamma\gamma X$ at the LHC, 
where this production rate can be determined with a 
relative error of 10-15\%\cite{test_hgaga_lhc}. 
Also, such a deviation in the branching ratio of 
$h \to \gamma\gamma$ directly affects the cross section 
of $e^+e^- \to  \nu\overline{\nu} h \;({\rm and}  Z^0 h)  
\to \nu\overline{\nu}  \gamma\gamma$, 
which can be measured with an accuracy of 16-22\%
at the future $e^+e^-$ LC (with $\sqrt{s}=500$ GeV and 
the integrated luminosity of 1 ab$^{-1}$)\cite{test_e+e-lc}.
Therefore, the Zee-model with low cut-off scales can be tested 
through the $h\to\gamma\gamma$ process at the LHC and the $e^+e^-$ LC's. 
At the future photon colliders, the enhancement (or reduction) 
of the $h\to\gamma\gamma$ partial decay rate will manifest itself 
in the different production rate of $h$ from the SM prediction. 
A few per cent of the deviation in 
$\Gamma(h\to\gamma\gamma) \cdot B(h \to b\overline{b})$ 
can be detected at a photon collider\cite{test_hgaga_lc}, so that the effects 
of the singlet charged Higgs boson can be tested even if the cutoff 
scale $\Lambda$ is at the Planck scale.   

The collider phenomenology of the singlet charged Higgs boson has turned 
out to be completely different from that of the THDM-like charged Higgs boson. 
The singlet charged Higgs boson mainly decays into $\ell^\pm \etv$ 
(with $\ell^\pm=e^\pm$ or $\mu^\pm$), while the decay mode $\tau^\pm \etv$ 
is almost negligible due to the relation  
$|f_{12}| \gg |f_{13}| \gg |f_{23}|$.  
This hierarchy among the coupling constants $f_{ij}$  
results from demanding bi-maximal mixings in the neutrino mass matrix 
generated in the Zee-model to be consistent with the neutrino 
oscillation data.  
On the other hand, the THDM-like charged Higgs boson decays mainly 
into either the $\tau \nu$ mode or the $c s$ mode,  
through the usual Yukawa-interactions.   
Hence, to probe this singlet charged Higgs boson using the LEP-II data,  
experimentalists should examine their data sample with 
$e^+e^-\etv$, $e^+\mu^-\etv$, $\mu^+e^-\etv$ or $\mu^+\mu^-\etv$, 
while the experimental lower mass bound of the THDM-like charged Higgs 
boson is obtained from examining the 
$\tau \tau \etv$, $\tau \etv jj$ and $jjjj$ events. 
Using the published LEP-II constraints on the MSSM smuon production 
(assuming the lightest neutralinos to be massless), 
we estimate the current lower mass bound of this singlet charged 
Higgs boson to be about 80-85 GeV. 
The Tevatron Run-II, LHC and future LC's can further test this model.

Finally, we comment on a case in which the singlet charged Higgs boson 
($S_2^\pm$ for $\chi=0$) is the lightest of all the Higgs bosons.  
For $m_h/2 >  m_{S_2} >  m_Z$, the Higgs sector of the Zee-model 
can be further tested by measuring the production rate of 
$pp \; ({\rm or}\; p\overline{p}) 
 \to h X \to S_2^+S_2^- X \to \ell^+\ell'^- \etv X$. 
The branching ratio for $h \ra S^+_2 S^-_2 \ra \ell^+ {\ell'}^- \etv$ 
can be large. For instance, for $m_h^{} = 210$ GeV and $m_{S_2^{}}^{}=100$ 
GeV, this branching ratio is about 12\% for each 
$\ell^+ {\ell'}^- = e^+e^-$, $e^+\mu^-$, $\mu^+e^-$ or $\mu^+\mu^-$. 
The branching ratio decreases for larger masses of $h$.
Moreover, the total decay width of $h$ can be largely 
modified when the decay channel $h \to S_2^+S_2^-$ is open. 
In this case, the decay branching ratios of $h \to W^+W^-$, $ZZ$ are 
also different from the SM predictions.  

In conclusion, the distinguishable features of the Zee-model from 
the SM and the THDM can be tested by the data from LEP-II, the Tevatron Run-II 
and future experiments at LHC and LC's.

\acknowledgments

We are grateful to the warm hospitality of the
Center for Theoretical Sciences in Taiwan where part of this
work was completed. CPY would like to thank H.-J. He, J. Ng and W. Repko
for stimulating discussions.
SK was supported, in part, by the Alexander von Humboldt Foundation.
GLL and JJT were supported, in part, by the National Science Council of 
R.O.C. under the Grant No NSC-89-2112-M-009-035;
YO  was supported by the Grant-in-Aid of the
Ministry of Education, Science, Sports and Culture, Government of Japan
(No.\ 09640381), Priority area ``Supersymmetry and Unified Theory of
Elementary Particles'' (No.\ 707), and ``Physics of CP Violation''
(No.\ 09246105); CPY is supported by
the National Science Foundation in the USA under the grant PHY-9802564.

\newpage
\appendix 
\section*{One-loop RGE's for dimensionless coupling constants}

Here, we summarize the relevant RGE's to our study. 
For the gauge coupling constants, we have 
\label{sec:rgelist}
\begin{eqnarray}
\mu \frac{d}{d\mu }g_{1} & = & \frac{1}{16\pi ^{2}}\frac{22}{3}g_{1}^{3},\\
\mu \frac{d}{d\mu }g_{2} & = & \frac{1}{16\pi ^{2}}\left( -3\right) 
g_{2}^{3},\\
\mu \frac{d}{d\mu }g_{3} & = & \frac{1}{16\pi ^{2}}\left( -7\right) g_{3}^{3}.
\end{eqnarray}
The RGE's for the Higgs-self-coupling constants of the doublets 
are calculated at the one-loop level as  
\begin{eqnarray}
\mu \frac{d}{d\mu }\lambda _{1} & = & 
\frac{1}{16\pi ^{2}}\left\{ 12\lambda _{1}^{2}+4\lambda _{3}^{2}+
4\lambda _{3}\lambda _{4}+2\lambda _{4}^{2}+2\lambda _{5}^{2}+
2\sigma _{1}^{2}\right. \nonumber \\
 &  & \left. -\left( 3g_{1}^{2}+9g_{2}^{2}\right) \lambda _{1}+
 \left( \frac{3}{4}g_{1}^{4}+\frac{3}{2}g_{1}^{2}g_{2}^{2}+
 \frac{9}{4}g_{2}^{4}\right) \right\},  \\
\mu \frac{d}{d\mu }\lambda _{2} & = & 
\frac{1}{16\pi ^{2}}\left\{ 12\lambda _{2}^{2}+4\lambda _{3}^{2}+
4\lambda _{3}\lambda _{4}+2\lambda _{4}^{2}+2\lambda _{5}^{2}+
2\sigma _{2}^{2}+12y_{t}^{2}\lambda _{2}-12y_{t}^{4}\right. \nonumber \\
 &  & \left. -\left( 3g_{1}^{2}+9g_{2}^{2}\right) \lambda _{2}+
 \left( \frac{3}{4}g_{1}^{4}+\frac{3}{2}g_{1}^{2}g_{2}^{2}+
 \frac{9}{4}g_{2}^{4}\right) \right\} \label{rge_lambda2} , \\
\mu \frac{d}{d\mu }\lambda _{3} & = & 
\frac{1}{16\pi ^{2}}\left\{ 2\left( \lambda _{1}+
\lambda _{2}\right) \left( 3\lambda _{3}+\lambda _{4}\right) +
4\lambda _{3}^{2}+2\lambda _{4}^{2}+2\lambda _{5}^{2}+
2\sigma _{1}\sigma _{2}+6y_{t}^{2}\lambda _{3}\right. \nonumber \\
 &  & \left. -\left( 3g_{1}^{2}+9g_{2}^{2}\right) \lambda _{3}+
 \left( \frac{3}{4}g_{1}^{4}-\frac{3}{2}g_{1}^{2}g_{2}^{2}+
 \frac{9}{4}g_{2}^{4}\right) \right\},  \\
\mu \frac{d}{d\mu }\lambda _{4} & = & 
\frac{1}{16\pi ^{2}}\left\{ 2\left( \lambda _{1}+
\lambda _{2}\right) \lambda _{4}+4\left( 2\lambda _{3}+
\lambda _{4}\right) \lambda _{4}+8\lambda _{5}^{2}+
6y_{t}^{2}\lambda _{4}\right. \nonumber \\
 &  & \left. -\left( 3g_{1}^{2}+9g_{2}^{2}\right) \lambda _{4}+
 3g_{1}^{2}g_{2}^{2}\right\} , \\
\mu \frac{d}{d\mu }\lambda _{5} & = & 
\frac{1}{16\pi ^{2}}\left\{ 2\lambda _{1}+2\lambda _{2}+
8\lambda _{3}+12\lambda _{4}+6y_{t}^{2}
 -\left( 3g_{1}^{2}+9g_{2}^{2}\right) \right\}  \lambda _{5} ,
\end{eqnarray}
and those with respect to the additional singlet charged Higgs are given by  
\begin{eqnarray}
\mu \frac{d}{d\mu }\sigma _{1} & = & 
\frac{1}{16\pi ^{2}}\left\{ 4\sigma _{1}^{2}+2\sigma _{1}\sigma _{3}+
6\lambda _{1}\sigma _{1}+
\left( 4\lambda _{3}+2\lambda _{4}\right) \sigma _{2}+
8f_{ij}f_{ij}\sigma _{1}\right. \nonumber \\
 &  & \left. -\left( \frac{15}{2}g_{1}^{2}+
 \frac{9}{2}g_{2}^{2}\right) \sigma _{1}+3g_{1}^{4}\right\},  \\
\mu \frac{d}{d\mu }\sigma _{2} & = & 
\frac{1}{16\pi ^{2}}\left\{ 4\sigma _{2}^{2}+
2\sigma _{2}\sigma _{3}+6\lambda _{2}\sigma _{2}+
\left( 4\lambda _{3}+2\lambda _{4}\right) \sigma _{1}+
6y_{t}^{2}\sigma _{2}+8f_{ij}f_{ij}\sigma _{2}\right. \nonumber \\
 &  & \left. -\left( \frac{15}{2}g_{1}^{2}+
 \frac{9}{2}g_{2}^{2}\right) \sigma _{2}+3g_{1}^{4}\right\},  \\
\mu \frac{d}{d\mu }\sigma _{3} & = & 
\frac{1}{16\pi ^{2}}\left\{ 8\sigma _{1}^{2}+
8\sigma _{2}^{2}+5\sigma _{3}^{2}+
16f_{ij}f_{ij}\sigma _{3}-128\, tr\, f^{4} 
 -12g_{1}^{2}\sigma _{3}+24g_{1}^{4}\right\} .
\end{eqnarray}
Finally, the RGE's for the Yukawa-type coupling constants are obtained 
at one-loop level as  
\begin{eqnarray}
\mu \frac{d}{d\mu }y_{t} & = & 
\frac{1}{16\pi ^{2}}\left\{ -\left( \frac{17}{12}g_{1}^{2}+
\frac{9}{4}g_{2}^{2}+8g_{3}^{2}\right) y_{t}+
\frac{9}{2}y_{t}^{3}\right\}, \\
\mu \frac{d}{d\mu }f_{ij} & = & 
\frac{1}{16\pi ^{2}}\left\{ -\left( \frac{3}{2}g_{1}^{2}+
\frac{9}{2}g_{2}^{2}\right) f_{ij}+4f_{kl}f_{kl}f_{ij}-
4f_{ik}f_{kl}f_{lj}\right\},  
\end{eqnarray}
where 
\begin{eqnarray}
 tr\, f^{4}&\equiv& \sum _{i,j,k,l=1-3} f_{ij}f_{jk}f_{kl}f_{li}, \nonumber  \\
 f_{ij}f_{ij}&\equiv& \sum _{i,j=1-3}f_{ij}f_{ij}. \nonumber 
\end{eqnarray}

\end{document}